\documentclass[useAMS,usenatbib,usegraphicx]{mn2e}

\newcommand{\be}{\begin{equation}}
\newcommand{\ee}{\end{equation}} 

\newcommand{\half}{\mbox{$\frac{1}{2}$}}

\newcommand{\cmMMM}{\mbox{cm$^{-3}$}}
\newcommand{\kms}{\mbox{km s$^{-1}$}}
\newcommand{\yr}{\mbox{yr}}
\newcommand{\muG}{\mbox{$\mu$G}}

\newcommand{\ltsim}{\la}
\newcommand{\gtsim}{\ga}
\newcommand{\Order}{\mbox{${\cal O}$}} 

\newcommand{\goesto}{\mbox{~~$\longrightarrow$~~}}

\newcommand{\vecxi}{\mbox{\boldmath{$\xi$}}}
\newcommand{\vecxiz}{\mbox{$\vecxi_0$}}
\newcommand{\vecxiem}{\mbox{$\vecxi_m$}}
\newcommand{\vecxhat}{\mbox{$\hat{\rm\bf x}$}}
\newcommand{\vecyhat}{\mbox{$\hat{\rm\bf y}$}}
\newcommand{\veczhat}{\mbox{$\hat{\rm\bf z}$}}
\newcommand{\vecfhat}{\mbox{$\hat{\rm\bf f}$}}

\newcommand{\matQ}{\mbox{${\sf Q}$}}

\newcommand{\Alfven}{\mbox{Alfv\'{e}n}}

\newcommand{\sigvin}{\mbox{$\left<\sigma v\right>_{\rm in}$}}
\newcommand{\tauin}{\mbox{$\tau_{{\rm in}}$}}
\newcommand{\emi}{\mbox{$m_{\rm i}$}}
\newcommand{\eni}{\mbox{$n_{\rm i}$}}
\newcommand{\rhoi}{\mbox{$\rho_{\rm i}$}}
\newcommand{\vi}{\mbox{$v_{{\rm i}}$}}
\newcommand{\Ti}{\mbox{$T_{{\rm i}}$}}
\newcommand{\Te}{\mbox{$T_{{\rm e}}$}}

\newcommand{\tauni}{\mbox{$\tau_{{\rm ni}}$}}
\newcommand{\emn}{\mbox{$m_{\rm n}$}}
\newcommand{\enn}{\mbox{$n_{\rm n}$}}
\newcommand{\rhon}{\mbox{$\rho_{\rm n}$}}
\newcommand{\vn}{\mbox{$v_{{\rm n}}$}}
\newcommand{\Tn}{\mbox{$T_{{\rm n}}$}}

\newcommand{\rhoiz}{\mbox{$\rho_{{\rm i}0}$}}
\newcommand{\Bz}{\mbox{$B_0$}}
\newcommand{\viAz}{\mbox{$v_{\rm iA0}$}}
\newcommand{\rxt}{\mbox{$r\left(x,t\right)$}}
\newcommand{\uxt}{\mbox{$u\left(x,t\right)$}}
\newcommand{\un}{\mbox{$u_{\rm n}$}}
\newcommand{\unxt}{\mbox{$\un\left(x,t\right)$}}
\newcommand{\rdot}{\mbox{$\dot{r}$}}
\newcommand{\udot}{\mbox{$\dot{u}$}}
\newcommand{\uprime}{\mbox{$u^{\prime}$}}
\newcommand{\bdot}{\mbox{$\dot{b}$}}
\newcommand{\bprime}{\mbox{$b^{\prime}$}}

\newcommand{\eikx}{\mbox{$e^{\imath kx}$}}
\newcommand{\emikx}{\mbox{$e^{-\imath kx}$}}
\newcommand{\xt}{\mbox{$\left(x,t\right)$}}
\newcommand{\kt}{\mbox{$\left(k,t\right)$}}
\newcommand{\modk}{\mbox{$\left|k\right|$}}
\newcommand{\modx}{\mbox{$\left|x\right|$}}

\newcommand{\rz}{\mbox{$r_0$}}
\newcommand{\rh}{\mbox{$r_{\rm h}$}}
\newcommand{\rp}{\mbox{$r_{\rm p}$}}
\newcommand{\rhat}{\mbox{$\hat{r}$}}
\newcommand{\rhatz}{\mbox{$\rhat_0$}}
\newcommand{\rhath}{\mbox{$\hat{r}_{\rm h}$}}

\newcommand{\bz}{\mbox{$b_0$}}
\newcommand{\bh}{\mbox{$b_{\rm h}$}}
\newcommand{\bp}{\mbox{$b_{\rm p}$}}
\newcommand{\bhat}{\mbox{$\hat{b}$}}
\newcommand{\bhatz}{\mbox{$\bhat_0$}}
\newcommand{\bhath}{\mbox{$\hat{b}_{\rm h}$}}
\newcommand{\bhatp}{\mbox{$\hat{b}_{\rm p}$}}

\newcommand{\uz}{\mbox{$u_0$}}
\newcommand{\uh}{\mbox{$u_{\rm h}$}}
\newcommand{\up}{\mbox{$u_{\rm p}$}}
\newcommand{\uhat}{\mbox{$\hat{u}$}}
\newcommand{\uhatz}{\mbox{$\uhat_0$}}
\newcommand{\uhath}{\mbox{$\hat{u}_{\rm h}$}}
\newcommand{\uhatp}{\mbox{$\hat{u}_{\rm p}$}}
\newcommand{\unhat}{\mbox{$\hat{u}_{\rm n}$}}
\newcommand{\unhatkt}{\mbox{$\unhat\kt$}}

\newcommand{\vecyhatkt}{\mbox{$\vecyhat\kt$}}
\newcommand{\vecyhath}{\mbox{$\vecyhat_{\rm h}$}}
\newcommand{\vecyhatp}{\mbox{$\vecyhat_{\rm p}$}}
\newcommand{\vecfhatkt}{\mbox{$\vecfhat\kt$}}
\newcommand{\Cmn}{\mbox{$C_-$}}
\newcommand{\Cpl}{\mbox{$C_+$}}

\newcommand{\omegam}{\mbox{$\omega_-$}}
\newcommand{\omegaz}{\mbox{$\omega_0$}}
\newcommand{\omegap}{\mbox{$\omega_+$}}
\newcommand{\omegapm}{\mbox{$\omega_{\pm}$}}
\newcommand{\kc}{\mbox{$k_{\rm c}$}}

\newcommand{\Gp}{\mbox{$G_+$}}
\newcommand{\Gm}{\mbox{$G_-$}}
\newcommand{\Gprm}{\mbox{$G^{\prime}$}}
\newcommand{\Gdot}{\mbox{$\dot{G}$}}

\newcommand{\Ghat}{\mbox{$\hat{G}$}}
\newcommand{\Gphat}{\mbox{$\hat{G}_+$}}
\newcommand{\Gmhat}{\mbox{$\hat{G}_-$}}
\newcommand{\Gdothat}{\mbox{$\hat{\dot{G}}$}}
\newcommand{\Gprmhat}{\mbox{$\hat{G}^{\prime}$}}

\newcommand{\xprm}{\mbox{$x^{\prime}$}}
\newcommand{\zprm}{\mbox{$z^{\prime}$}}
\newcommand{\tprm}{\mbox{$t^{\prime}$}}

\newcommand{\ug}{\mbox{$u_{\rm g}$}}
\newcommand{\Dhat}{\mbox{$\hat{D}$}}
\newcommand{\Phat}{\mbox{$\hat{P}$}}
\newcommand{\Qhat}{\mbox{$\hat{Q}$}}
\newcommand{\Shat}{\mbox{$\hat{S}$}}
\newcommand{\Shatmn}{\mbox{$\hat{S}_-$}}
\newcommand{\Shatpl}{\mbox{$\hat{S}_+$}}
\newcommand{\emikugt}{\mbox{$e^{-\imath ku_{\rm g}t}$}}
\newcommand{\emiwmt}{\mbox{$e^{-\imath \omega_{-}t}$}}
\newcommand{\emiwpt}{\mbox{$e^{-\imath \omega_{+}t}$}}
\newcommand{\uhatps}{\mbox{$\hat{u}_{\rm ps}$}}
\newcommand{\uhatpt}{\mbox{$\hat{u}_{\rm pt}$}}
\newcommand{\ups}{\mbox{$u_{\rm ps}$}}
\newcommand{\upt}{\mbox{$u_{\rm pt}$}}
\newcommand{\Dtilde}{\mbox{$\tilde{D}$}}

\title[Driven waves in a two-fluid plasma]{Driven waves in a two-fluid plasma}
\author[W. G. Roberge and Glenn E. Ciolek]{W. G. Roberge\thanks{E-mail:
roberw@rpi.edu (WGR); cioleg@rpi.edu (GEC)} and Glenn E. Ciolek\\
Department of Physics, Applied Physics and Astronomy,
Rensselaer Polytechnic Institute,  
110 8th Street, Troy, NY 12180 USA}

\begin{document}
\maketitle

\begin{abstract}
We study the physics of wave propagation in a weakly ionised plasma,
as it applies to the formation of multifluid, MHD shock waves.
We model the plasma as separate charged and neutral fluids which
are coupled by ion-neutral friction.
At times much less than the ion-neutral drag time, the fluids are
decoupled and so evolve independently.
At later times, the evolution is determined by the large
inertial mismatch between the charged and neutral particles.
The neutral flow continues to evolve independently; the charged
flow is driven by and slaved to the neutral flow by friction.
We calculate this driven flow analytically by considering the special but
realistic case where the charged fluid obeys linearized equations of motion.
We carry out an extensive analysis of linear, driven, MHD waves.
The physics of driven MHD waves is embodied in certain
Green functions which describe wave propagation on short time scales,
ambipolar diffusion on long time scales, and transitional behavior
at intermediate times.
By way of illustration, we give an approximate solution for the
formation of a multifluid shock during the collision of two
identical interstellar clouds.
The collision produces forward- and reverse J shocks in the neutral
fluid and a transient in the charged fluid.
The latter rapidly evolves into a pair of magnetic precursors on
the J shocks, wherein the ions undergo force free motion and the
magnetic field grows monotonically with time.
The flow appears to be self similar at the time when linear
analysis ceases to be valid.
\end{abstract}


\section{Introduction}
\label{sec-intro}

It is well known that shock waves in weakly-ionised interstellar plasmas
have a multifluid structure, where the charged and neutral components
of the plasma behave as separate, interacting fluids (\citealt{Mullan71}).
Because their multifluid nature has profound observational consequences
(\citealt{Draine80}), multifluid shocks have been the subject of
numerous studies on their structure, chemistry, and emission
(last reviewed by \citealt{DM93}).
Although the majority of these studies have assumed steady flow,
a small fraction have carried out time dependent simulations,
either to follow the development of instabilities
(\citealt{MS97}; \citealt{Stone97}; \citealt{NS97})
or to study evolutionary effects
(\citealt{SM97}; Chi\`{e}ze, Pineau des For\^{e}ts \& Flower 1998;
\citealt{CR02}; Lesaffre et al.\ 2004a,b).

This paper describes the formation of a
multifluid shock wave by a sudden disturbance, e.g., the collision of two
cloud cores or the impact of a protostellar outflow onto surrounding core material.
We focus on time scales $< 100\,\tauin$, where the ion-neutral
drag time, \tauin, is the slowing-down time for an ion drifting
through a neutral gas.
This is a {\it very}\/ short time:
$\tauin \sim 0.01$\,yr in a typical dense core (\S\ref{sec-governing}).
Nevertheless, there are good reasons for studying this extremely brief,
unobservable period.
First, the physics is interesting.
The response of a plasma to a sudden disturbance
depends on the wave modes it supports over a broad range in frequency, $\nu$.
In a weakly ionised plasma the allowed modes represent propagating waves
if $\nu \gtsim\tau_{\rm in}^{-1}$ and diffusion if
$\nu \ltsim\tau_{\rm in}^{-1}$ (\S\ref{sec-dispersion}).
We wish to understand how these qualitatively different behaviors
manifest themselves at early times, and whether the resulting
effects influence the flow at later, observable times.
Second, the present study serves as a prototype for future work,
on the effects of charged dust grains on multifluid shocks.
These effects are known to be important
(\citealt{Wardle98}; \citealt{CR02};
Ciolek, Roberge \& Mouschovias 2004; \citealt{CW06}).
Including dust will be analogous in some ways to the present study,
but dust also adds new physics and a higher level of complexity to the problem.
Third, the results of this paper have practical use.
As a computational expedient, some time dependent simulations
of multifluid shock waves neglect the inertia of the charged fluid
(e.g., \citealt{SM97}).
Since the inertia is important on precisely the time scales studied here,
the present work provides benchmark tests on this assumption.

The plan of this paper is as follows.
In \S\ref{sec-governing} we give the equations of motion for
a time dependent, multifluid shock wave in a form which exploits
the small time scale of interest.
In \S\ref{sec-drivenwaves} the linearized versions of
these equations (\S\ref{sec-linearization}) are presented and we
discuss the allowed wave modes 
(\S\S\ref{sec-fourier}--\ref{sec-dispersion}).
We emphasize that linear analysis yields highly accurate solutions
for some special but realistic cases; the general case will be
studied numerically in a separate paper. Analytical methods for 
calculating the time dependent flow of charged and neutral particles are
described in \S\S\ref{sec-yhsol}--\ref{sec-steady}. These techniques 
are used in \S\ref{sec-collision} to find an approximate solution for 
the formation of a multifluid shock during the collision of two 
identical clouds. Our results are summarized in \S\ref{sec-summary}.

\section{Governing Equations}
\label{sec-governing}

We are interested in the multifluid flow which ensues when a weakly 
ionized plasma is accelerated and compressed by a sudden disturbance. 
We model the plasma as separate charged and neutral fluids which
interact via elastic scattering.
The charged fluid is composed of ions and electrons (the effects
of charged dust grains will be considered in a separate paper)
plus a magnetic field which is everywhere frozen into the charged
fluid. We adopt planar geometry with the magnetic field along the $z$ 
direction and fluid velocities along the $\pm x$ directions.

The equations of motion for an arbitrary disturbance in a two-fluid
plasma were derived by Draine 
(1986; see also Nemirovsky, Fredkin \& Ron 2002).
Here we solve modified versions of these equations which take advantage
of the extremely short time scale--- typically less than $1$\,\yr--- 
on which we follow the flow.
The charged fluid is described by the equations of mass conservation,
momentum conservation, and the induction equation:
\be
\frac{\partial\rho_{\rm i}}{\partial t} + 
\frac{\partial}{\partial x}\left(\rho_{\rm i} v_{\rm i}\right) = 0,
\label{eq-mass}
\ee
\be
\frac{\partial v_{\rm i}}{\partial t}
+
v_{\rm i}\frac{\partial v_{\rm i}}{\partial x}
=
\frac{v_{\rm n}-v_{\rm i}}{\tau_{\rm in}}
- \frac{1}{\rho_{\rm i}}\frac{\partial}{\partial x}\left(\frac{B^2}{8\pi}\right),
\label{eq-momentum}
\ee
and
\be
\frac{\partial B}{\partial t} + 
\frac{\partial}{\partial x}\left(B v_{\rm i}\right) = 0
\label{eq-induction}
\ee
respectively,
where $\rhoi$ and $\vi$ are the density and
velocity of the charged fluid,
\vn\ is the velocity of the neutral fluid,
and $B$ is the magnetic field.
The first term on the RHS of eq.~(\ref{eq-momentum}) is
the frictional acceleration produced by
elastic scattering between ions and neutral particles
and the second is the acceleration caused by the magnetic pressure gradient.

The characteristic time scale for acceleration by friction is the 
ion-neutral drag time, \tauin. If the charged and neutral fluids were
each composed of a single species, then
\be
\tauin = \frac{1+\emi/\emn}{n_{\rm n}\left<\sigma v\right>_{\rm in}},
\label{eq-tauin}
\ee
where \emn\ and \emi\ are the neutral and ion particle masses, respectively,
\enn\ is the number density of the neutral fluid,
and \sigvin\ is the momentum transfer rate coefficient for elastic 
ion-neutral scattering.
For a typical cloud core with $\enn=2 \times 10^4$\,\cmMMM\ and $\emi=25\emn$,
one finds $\tauin\sim 0.01$\,yr.
%
%
%
%
This is a fundamental time scale for the flow.

We have omitted the energy equation for the charged fluid because
eq.~(\ref{eq-mass})--(\ref{eq-momentum}) are independent of the
ion and electron temperatures, \Ti\ and \Te.
This is appropriate at the very low fractional ionizations
$\left(\ltsim 10^{-8}\right)$ of interest here, where the pressure
of the charged fluid is dominated by magnetic pressure.
The cooling rate of the neutral fluid generally
depends on \Ti\ and \Te\ (via the rates of ion- and electron impact 
processes) but we assume that radiative cooling is negligible
on the time scales of interest.
For a gas with $\enn=2\times 10^4$\,\cmMMM, the cooling time is
$>1$\,yr if the neutral temperature is less than $\approx 2300$\,K.
In eq.~(\ref{eq-mass}) we have omitted a term
which represents mass transfer between the charged and neutral fluids.
This is always a good approximation because the recombination time scale is
always $\gg 1$\,\yr\ in dense clouds.

We assume that the neutral fluid is governed by Euler's equations
for adiabatic flow,
\be
\frac{\partial\rho_{\rm n}}{\partial t}
+
\frac{\partial}{\partial x}
\left( \rhon v_{\rm n} \right)
=0,
\label{eq-nmass}
\ee
\be
\frac{\partial v_{\rm n}}{\partial t}
+
\vn\,\frac{\partial v_{\rm n}}{\partial x}
=
-
\frac{1}{\rho_{\rm n}}\,\frac{\partial P_{\rm n}}{\partial x},
\label{eq-nmomentum}
\ee
and
\be
P_{\rm n} = K\,\rho_{\rm n}^{\gamma}.
\ee
Mass transfer in eq.~(\ref{eq-nmass}) is also neglected, for reasons
noted above.
We have neglected momentum transfer by friction in eq.~(\ref{eq-nmomentum})
because the time for friction to accelerate the neutral fluid is
\be
\tauni = \frac{\rho_n}{\rho_i}\,\tauin,
\ee
or about $10^4$\,yr\ in a dense
core.\footnote{Of course this means that the total momentum of the
charged plus neutral fluids is not conserved but the associated
error is $\Order\left(\tauin/\tauni\right) \sim 10^{-6}$.}
We have also neglected heating of the neutral fluid by ion-neutral
friction and the associated acceleration by thermal pressure gradients;
one can show that these effects are small on
time scales $\sim 1$\,yr.

The equations of motion for the charged fluid depend on the
neutral velocity, \vn, and number density, \enn.
In the next section, on the dynamics of the charged fluid,
we assume that \vn\ and \enn\ are known functions of
$x$ and $t$ which have been determined by solving Euler's equations.
How this all works out for a particular example is
demonstrated in \S\ref{sec-collision}.

\section{The Physics of Driven Waves}
\label{sec-drivenwaves}

\subsection{Linearized Equations of Motion}
\label{sec-linearization}

We follow the flow of the charged fluid by solving
the {\it linearized}\/ versions of  eq.~(\ref{eq-mass})--(\ref{eq-induction}).
This is an expedient which allows us to work the problem analytically.
Our analytical solutions reveal the essential physics;
numerical solutions of the full nonlinear equations will be discussed elsewhere.
However it is important to note that linear theory is highly accurate
for some initial conditions at sufficiently short times; this is
ultimately because the speed of a typical disturbance ($\sim 10$--$100$\,\kms)
is much smaller than the ion \Alfven\ speed ($\sim 1000$\,\kms).
A specific example is discussed in \S\ref{sec-collision}.

The zero-order solution of eq.~(\ref{eq-mass})--(\ref{eq-induction})
is a spatially uniform state in which the charged fluid has
constant density \rhoiz, magnetic field, \Bz, and zero velocity.
Onto the zero-order state we superpose density, magnetic field, and
velocity perturbations denoted $r$, $b$, and $u$, respectively.
We adopt dimensionless perturbations so that
\be
\rhoi\xt \equiv \rhoiz\,\left[1+\rxt\right],
\label{eq-rdef}
\ee
\be
B\xt \equiv \Bz\,\left[1+b\xt\right],
\label{eq-bdef}
\ee
and
\be
\vi\xt \equiv \viAz\,\uxt,
\label{eq-udef}
\ee
where
\be
\viAz \equiv \frac{B_0}{\sqrt{4\pi\rhoiz}}
\ee
is the ion \Alfven\ speed in the zero-order state.
%
%
%
%
We also adopt dimensionless time and distance units.
In all subsequent discussion, $t$ is dimensionless time in
units of \tauin\ and $x$ is dimensionless distance in units of
$\viAz\tauin$.
In this paper we shall assume that $\tauin$ is independent of $x$ and $t$.
This allows us to obtain analytical solutions but,
because \tauin\ depends on the density of the neutral fluid,
only hypothetical flows with uniform density can be studied
analytically.

Linearizing eqs.~(\ref{eq-mass})--(\ref{eq-induction}) about
the zero-order solution gives the equations of motion:
\be
\rdot ~+~ \uprime ~=~ 0,
\label{eq-mass-lin}
\ee
\be
\bdot ~+~ \uprime ~=~ 0,
\label{eq-induction-lin}
\ee
and
\be
\udot ~+~ u ~+~ \bprime ~=~\un,
\label{eq-momentum-lin}
\ee
where
\be
\unxt\equiv \vn\xt/\viAz
\ee
and the dots and primes denote partial derivatives with respect to
$t$ and $x$, respectively.
We seek solutions of equations (\ref{eq-mass-lin})--(\ref{eq-momentum-lin})
subject to the initial conditions
\be
r(x,0) = r_0(x),
\label{eq-initc-r}
\ee
\be
b(x,0) = b_0(x),
\label{eq-initc-b}
\ee
and
\be
u(x,0) = u_0(x),
\label{eq-initc-u}
\ee
where $r_0$, $b_0$, and $u_0$ are small but otherwise
arbitrary functions.
We assume that $u_{\rm n}$ has been determined as described in \S\ref{sec-governing}.
For the purposes of this section it is a known, small, but otherwise
arbitrary function.

%
%
%
%

\subsection{Fourier Analysis}
\label{sec-fourier}

To proceed we Fourier transform eq.~(\ref{eq-mass-lin})--(\ref{eq-momentum-lin})
to eliminate the spatial derivatives.
We define the Fourier transform of $f\xt$ by
\be
\hat{f}(k,t) \equiv \frac{1}{2\pi}\,\int_{-\infty}^{+\infty}\ dx\,\emikx\,f(x,t).
\label{eq-FTdef}
\ee
Writing the perturbations as Fourier integrals transforms
eq.~(\ref{eq-mass-lin})--(\ref{eq-momentum-lin}) into a set
of linear, inhomogeneous, coupled, ODEs for the transforms of the
perturbations:
\be
\frac{d\vecyhat}{dt}
~=~
\matQ\,\vecyhat ~+~ \vecfhat,
\label{eq-odeyhat}
\ee
where the vector of unknowns is
\be
\vecyhatkt \equiv \left[ \rhat\kt, \bhat\kt, \uhat\kt \right]^t,
\label{eq-yhatcomponents}
\ee
the coupling matrix is
\be
\matQ(k) ~\equiv~
\left(
\begin{array}{ccc}
  0  &  0        &  -\imath k \\
  0  &  0        &  -\imath k \\
  0  & -\imath k &  -1        \\
\end{array}
\right),
\label{eq-Qcomponents}
\ee
and the source term,
\be
\vecfhatkt \equiv \left[ 0, 0, \unhatkt \right]^t,
\label{eq-fhatcomponents}
\ee
represents the effects of frictional driving.

Equation (\ref{eq-odeyhat}) can be solved by standard methods.
The solution has the form
\be
\vecyhatkt = \vecyhath\kt ~+~ \vecyhatp\kt,
\ee
where the particular solution, \vecyhatp, is any solution of
eq.~(\ref{eq-odeyhat}) and the homogeneous solution, \vecyhath, is
any solution with $\vecfhat=0$.
The ``total'' solution must satisfy the initial conditions
\be
\vecyhatp(k,0)+\vecyhath(k,0) = \left[\rhatz,\bhatz,\uhatz\right]^t
\label{eq-totalic}
\ee
but there is a degree of arbitrariness in how the right side
of (\ref{eq-totalic}) is apportioned between the particular
and homogeneous parts on the left.
We require
\be
\vecyhatp(k,0) = 0
\label{eq-yhatp-ic}
\ee
and
\be
\vecyhath(k,0) = \left[\rhatz,\bhatz,\uhatz\right]^t.
\label{eq-yhath-ic}
\ee
Then for a given set of initial conditions,
the homogeneous solution represents the disturbance that would occur,
for the same initial conditions, if the driving force was zero.
%
%
%
%
We show in \S\ref{sec-dispersion} that the homogeneous solution has
no growing modes; it is therefore entirely transient in nature.
The particular solution always has a transient component
but may also contain a steady part if the driving is steady
(see \S\ref{sec-steady}).

\subsection{Dispersion Relations}
\label{sec-dispersion}

The solution of eq.~(\ref{eq-odeyhat}) depends on the eigenvalues and
eigenvectors of the coupling matrix.
It is easy to show that \matQ\ has three eigenvalues,
$-\imath\omegam$, $-\imath\omegaz$, and $-\imath\omegap$, where
\be
\omegaz = 0,
\label{eq-omegazdef}
\ee
and
\be
\omega_{\pm}(k) = -\frac{\imath}{2} \pm R(k).
\label{eq-omegapmdef}
\ee
The corresponding eigenvectors are
\be
\vecxiz = \left[1, 0, 0\right]^t
\ee
and
\be
\vecxi_{\pm} = \left[k/\omega_{\pm}, k/\omega_{\pm}, 1\right]^t,
\ee
respectively.

\begin{figure}
\includegraphics[width=77mm]{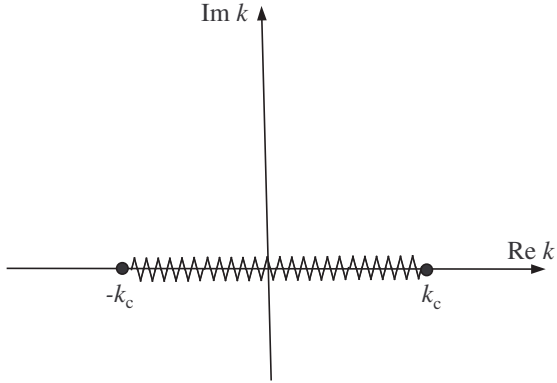}
\caption{Function $R(k)$ has a branch
cut in the complex $k$ plane along the interval
indicated by the sawtooth curve.}
\label{fig-branch}
\end{figure}
\begin{figure}
\includegraphics[width=65mm]{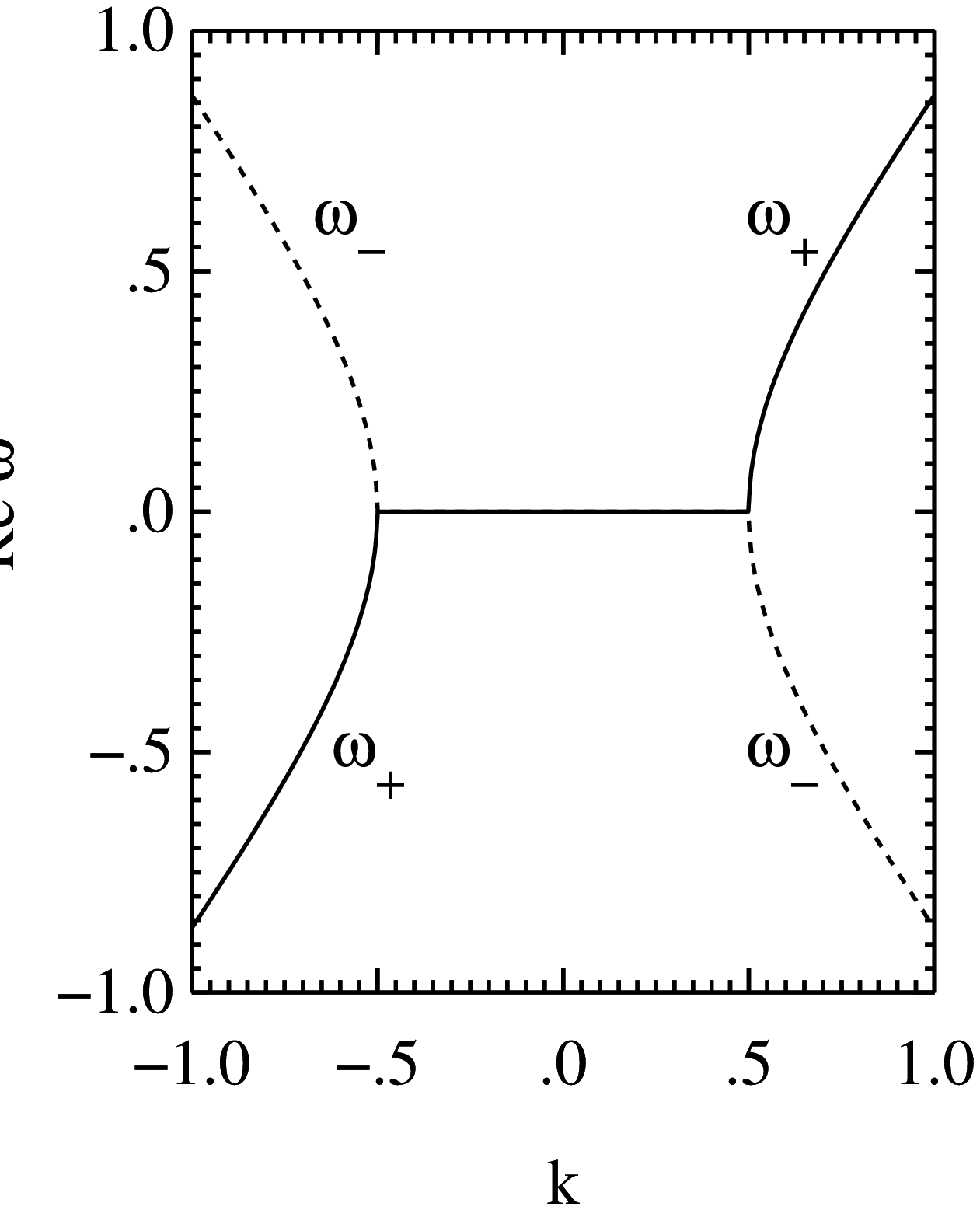}
\caption{Real part of \omegap\ (solid) and
\omegam\ (dashed) plotted vs.\ wave number $k$.
%
%
%
%
The curves are degenerate for $\left|k\right| \le 1/2$.
All quantities are dimensionless.}
\label{fig-dispr}
\includegraphics[width=65mm]{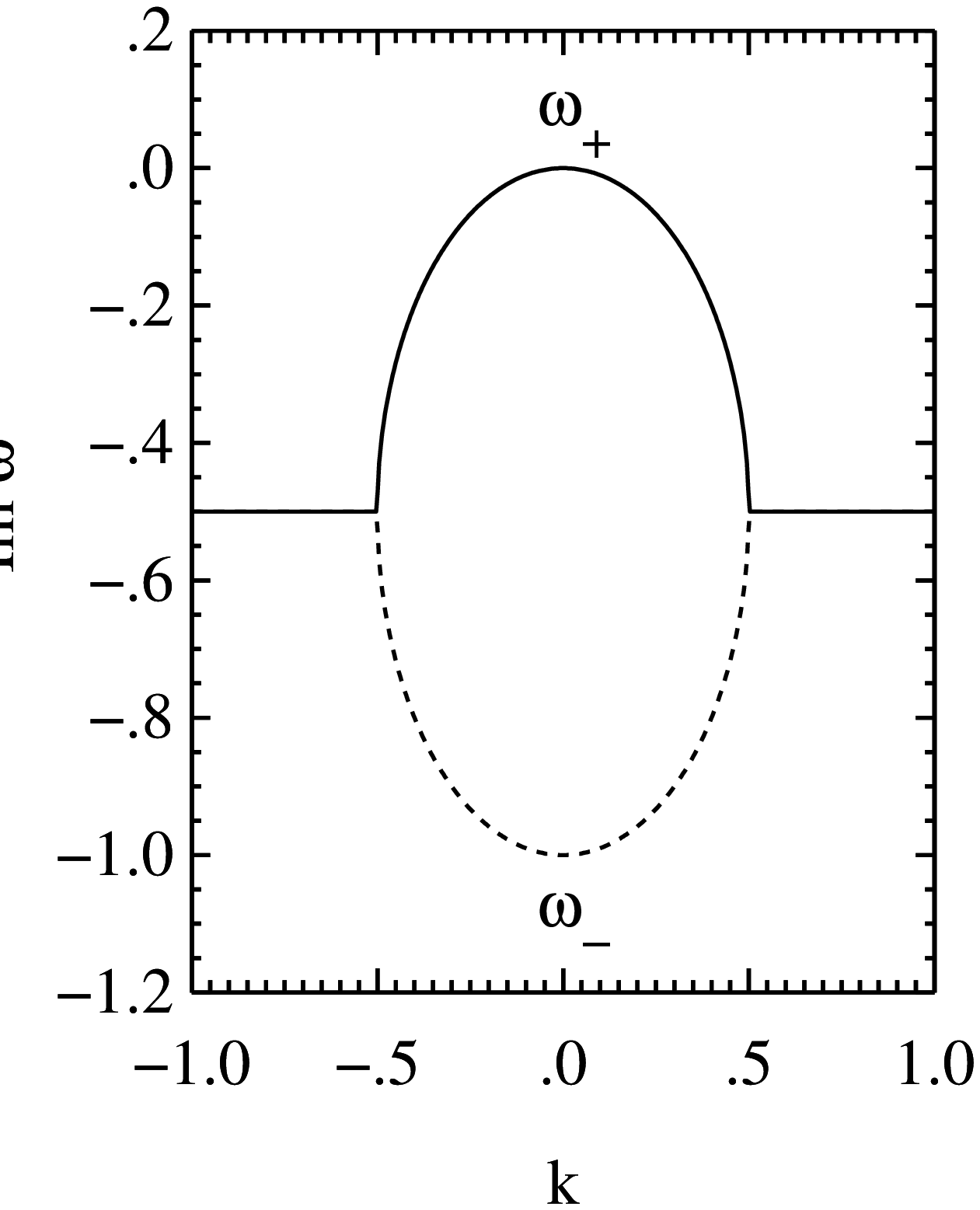}
\caption{Imaginary part of \omegap\ (solid) and
\omegam\ (dashed) plotted vs.\ wave number $k$.
%
%
%
%
The curves are degenerate for $\left|k\right| \ge 1/2$.
All quantities are dimensionless.}
\label{fig-dispi}
\end{figure}

The eigenvalues and eigenvectors depend on the complex-valued function
\be
R(k) \equiv \left(k^2-k_{\rm c}^2\right)^{1/2},
\label{eq-Rdef}
\ee
where $\kc=1/2$ is the dimensionless\footnote{In ordinary units,
$\kc=2\pi/\lambda_{\rm c}$, where
$\lambda_{\rm c}=4\pi\viAz\tauin$ is the maximum wavelength
for propagating ion magnetosound (and \Alfven) waves in a cold plasma
(\citealt{KP69}; \citealt{CRM04}).}
 ``critical wave number.''
For real $k$ the meaning of $R$ is ambiguous on the interval
$-\kc < k < +\kc$, where $R$ has a branch cut in the complex $k$ plane
(Fig.~\ref{fig-branch}).
We will take
\be
R(k) ~\equiv~ \left\{
\begin{array}{rc}
-\sqrt{k^2-k_{\rm c}^2}       &  k<-\kc    \\
                              &            \\
\imath\sqrt{k_{\rm c}^2-k^2}  & -\kc<k<\kc \\
                              &            \\
+\sqrt{k^2-k_{\rm c}^2}       &  k>+\kc    \\
\end{array}
\right.,
\label{eq-Rbranch}
\ee
where ``$\sqrt{~~~~}$'' denotes the positive square root.
This means that for real $k$ we evaluate $R(k)$ ``just above'' the 
branch cut, a fact which becomes crucial when Fourier transforms are evaluated
as contour integrals (App.~\ref{app-green}).
With our definition of $R$, the dependence of $\omegam$
and $\omegap$ on $k$ (for real $k$) is as shown in
Fig.~\ref{fig-dispr}--\ref{fig-dispi}.
Kulsrud and Pearce (1969) found the analogous dispersion relations for
\Alfven\ waves in a cold ion-neutral plasma.
%
%
%
%
For dimensionless wave numbers $k \gg \sqrt{\rhoi/\rhon}\sim 10^{-3}$,
where the neutrals act as a stationary background, their dispersion
relation reduces to expression (\ref{eq-omegapmdef}) for \omegapm\ 
[cf.\ \citealt{KP69}, eq.~(C7)].

%
The physics of the different wave modes
is well understood but worth repeating for later discussion.
The $\omegaz$ mode
represents perturbations (e.g., perturbations in
$B/\rhoi$) that move along with the charged fluid.
For $\left|k\right|>\kc$ the \omegap\ and \omegam\ modes
represent propagating waves (Fig.~\ref{fig-dispr}).
For very large wave numbers, the dispersion relations become
\be
\lim_{|k| \gg 1}
\omegapm = -\frac{i}{2} \pm k.
\ee
In this limit the phase velocity approaches
$\pm 1$ ($\leftrightarrow\pm\viAz$ in ordinary units) and the
damping time approaches $2$ ($\leftrightarrow2\tauin$).

For $\left|k\right|<\kc$ the \omegap\ and \omegam\ modes
are both evanescent (Fig.~\ref{fig-dispr}).
However they describe qualitatively different behavior.
For very small wave numbers the dispersion relations become
\be
\lim_{|k| \ll 1} \omegap(k) = -\imath k^2.
\label{eq-omegap-long}
\ee
and
\be
\lim_{|k| \ll 1} \omegam(k) = -\imath+\imath k^2.
\label{eq-omegam-long}
\ee
These are to be compared with the dispersion relation
\be
\omega_{\rm diff}(k) = -\imath k^2\alpha
\ee
for the diffusion equation with diffusion coefficient $\alpha$.
%
%
%
Evidently $\omegap$ represents diffusion with $\alpha=1$ 
$\left(\leftrightarrow v_{\rm iA0}^2\tauin\right)$ and
zero damping.
This ``diffusion mode'' dominates all solutions at large times.
%
%
%
%
In contrast, \omegam\ represents ``antidiffusion'' with $\alpha=-1$
and strong damping.
The \omegam\ mode describes the transient compression of $B$
which can occur at early times for some initial conditions.
However it is never important for $t\gtsim 1$.

\subsection{Homogeneous Solution}
\label{sec-yhsol}

The homogeneous solution of eq.~(\ref{eq-odeyhat}) is
\be
\vecyhath\kt = A_m\,e^{-\imath\omega_m t}\,\vecxiem,
\label{eq-yhath}
\ee
where summation is implied by the repeated index.
The coefficients $A_m$ are functions of $k$ but not time 
and are determined by the initial conditions.
Setting $t=0$ in eq.~(\ref{eq-yhath}) and substituting the
result into the LHS of eq.~(\ref{eq-yhath-ic}) gives a set of linear
algebraic equations for $\left\{A_m\right\}$.
The solution is
\be
A_- = \frac{\omega_-}{2R}
\left(\frac{\omega_+}{k}\,\bhatz-\uhatz\right)
\label{eq-Amsol}
\ee
\be
A_0 = \rhatz-\bhatz,
\label{eq-Azsol}
\ee
and
\be
A_+ = -\frac{\omega_+}{2R}
\left(\frac{\omega_-}{k}\,\bhatz-\uhatz\right).
\label{eq-Apsol}
\ee
The expansion coefficients obey the symmetry requirement
$A_- \rightarrow A_+$ under the interchange $- \rightarrow +$.
To see this, note that 
$R = (\omega_+-\omega_-)/2$ [cf.\ eq.~(\ref{eq-omegapmdef})].

The transforms of the perturbations are obtained by
substituting the expansion coefficients into eq.~(\ref{eq-yhath}).
%
%
After lengthy algebra we find that
\be
\rhath\kt = \bhath + \rhatz - \bhatz,
\label{eq-rhathsol}
\ee
\be
\bhath\kt =
2\pi\hat{G}\,\bhatz
+2\pi\Gdothat\,\bhatz
-2\pi\Gprmhat\,\uhatz,
\label{eq-bhathsol}
\ee
and
\be
\uhath\kt = 
 2\pi\Gdothat\,\uhatz
-2\pi\Gprmhat\,\bhatz.
\label{eq-uhathsol}
\ee
We have introduced the function
\be
\Ghat\kt \equiv \Gphat\kt - \Gmhat\kt,
\label{eq-ghatdef}
\ee
where
\be
\Gphat\kt \equiv \frac{\imath}{4\pi}\,\frac{e^{-\imath\omega_+ t}}{R},
\label{eq-gphatdef}
\ee
and
\be
\Gmhat\kt \equiv \frac{\imath}{4\pi}\,\frac{e^{-\imath\omega_- t}}{R}~;
\label{eq-gmhatdef}
\ee
$\Gprmhat\kt$ and $\Gdothat\kt$
are the Fourier transforms of
$G^{\prime}\xt$ and $\Gdot\xt$, respectively. 
We shall see shortly that \Ghat, \Gprmhat, and \Gdothat\ are the Fourier
transforms of certain Green functions.
The Green functions 
are calculated in Appendix~{\ref{app-green} 
and discussed in \S\ref{sec-green}.
Here we simply note that they are {\em causal}:
\be
G\xt = \Gdot\xt = \Gprm\xt = 0 ~~~{\rm if}~\left|x\right|>t.
\ee

The homogeneous solution for the perturbations is obtained by taking the
inverse Fourier transforms of expressions
(\ref{eq-rhathsol})--(\ref{eq-uhathsol}).
This can be done by inspection using the Convolution Theorem,
\be
\int_{-\infty}^{+\infty}\ dk\,\eikx\,\hat{f}(k)\,\hat{g}\kt
= \frac{1}{2\pi}
\int_{-\infty}^{+\infty}\ d\xprm\,f\left(\xprm\right)\,g\left(x-\xprm,t\right).
\ee
The result is the homogeneous solution:
\be
\rh\xt = \bh\xt+\rz(x)-\bz(x)
\label{eq-rhsol}
\ee
\be
\bh\xt = \left<G\left|\bz\right.\right>
+ \left<\Gdot\left|\bz\right.\right>
- \left<\Gprm\left|\uz\right.\right>
\label{eq-bhsol}
\ee
and
\be
\uh\xt = \left<\Gdot\left|\uz\right.\right>
          -\left<\Gprm\left|\bz\right.\right>,
\label{eq-uhsol}
\ee
where the angle brackets signify convolution,
\be
\left<g\left|f\right.\right> \equiv
\int_{x-t}^{x+t}\ d\xprm\,
g\left(x-\xprm,t\right)\,
f\left(\xprm\right),
\label{eq-convdef}
\ee
and causality has been used to refine the limits of integration.

Expression (\ref{eq-rhsol}) for the density perturbation
has a simple physical interpretation.
Taking the ratio of eq.~(\ref{eq-bdef}) and (\ref{eq-rdef}) and
neglecting terms of second order gives
\be
\frac{B\xt}{\rhoi\xt} = \frac{\Bz}{\rhoiz}\left[1+b\xt-\rxt\right].
\ee
But according to eq.~(\ref{eq-rhsol}),
the factor in square brackets is a conserved quantity so
\be
\frac{B\xt}{\rhoi\xt} = \frac{B(x,0)}{\rhoi(x,0)}.
\label{eq-flxfrz}
\ee
Noting that $\vi=0$ in the unperturbed state, we see that
expression (\ref{eq-flxfrz}) implies flux freezing.
Of course flux freezing was put into the solution {\em a priori},
so there was really no need to calculate the density perturbation
independently of the magnetic field perturbation.
However it is reassuring to know
that the homogeneous solution is self consistent in this respect.

\subsection{Particular Solution}
\label{sec-ypsol}

The particular solution of eq.~(\ref{eq-odeyhat}) has the form
\be
\vecyhatp\kt = C_m\,e^{-\imath\omega_m t}\,\vecxiem,
\label{eq-yhatpdef}
\ee
where the coefficients $C_m$ generally depend on time
as well as $k$.
One can verify that expression (\ref{eq-yhatpdef}) is a solution
of eq.~(\ref{eq-odeyhat}) provided
\be
\dot{C}_m\,e^{-\imath\omega_m t}\,\vecxiem = \vecfhat\kt.
\label{eq-odeCm}
\ee
Expression (\ref{eq-odeCm}) is a set of
coupled ODEs for $\left\{C_m\right\}$ with initial
conditions
\be
C_m(k,0)=0.
\label{eq-Cminitc}
\ee
The solution is
\be
C_-\kt = -\frac{\omega_-}{2R}\int_0^t\ d\tprm\ 
\unhat\left(k,\tprm\right)\,e^{\imath\omega_- t^{\prime}},
\label{eq-Cmsolution}
\ee
\be
C_0\kt = 0,
\ee
and
\be
C_+\kt = \frac{\omega_+}{2R}\int_0^t\ d\tprm\ 
\unhat\left(k,\tprm\right)\,e^{\imath\omega_+ t^{\prime}}.
\label{eq-Cpsolution}
\ee

Substituting the expansion coefficients into expression
(\ref{eq-yhatpdef}) yields the Fourier transforms of
the perturbations.
The inverse transform then yields the particular solution:
\be
\rp\xt=\bp\xt,
\label{eq-rpsol}
\ee
\begin{eqnarray}
\bp\xt &=& -\int_0^t d\tprm
              \int_{x-t+t^{\prime}}^{x+t-t^{\prime}} d\xprm\ 
              u_n\left(x^{\prime},t^{\prime}\right) \nonumber\\
& &     \times~ G^{\prime}\left(x-x^{\prime},t-t^{\prime}\right),\\
\label{eq-bpsol}
\nonumber
\end{eqnarray}
and
\begin{eqnarray}
\up\xt &=& \int_0^t d\tprm
           \int_{x-t+t^{\prime}}^{x+t-t^{\prime}} d\xprm\ 
           u_n\left(x^{\prime},t^{\prime}\right)  \nonumber\\
& &        \times~ \dot{G}\left(x-x^{\prime},t-t^{\prime}\right),\
            \label{eq-upsol} \\
\nonumber
\end{eqnarray}
where the Convolution Theorem has been used again.
Comparing eq.~(\ref{eq-rpsol}) to eq.~(\ref{eq-rhsol})
and noting that the particular solution vanishes at $t=0$,
we see that the particular solution also incorporates
flux freezing.
Since flux freezing uniquely relates the density perturbation
to the magnetic field perturbation, we omit further discussion of $r\xt$.

\subsection{Green Functions}
\label{sec-green}

\begin{figure}
\includegraphics[width=68mm]{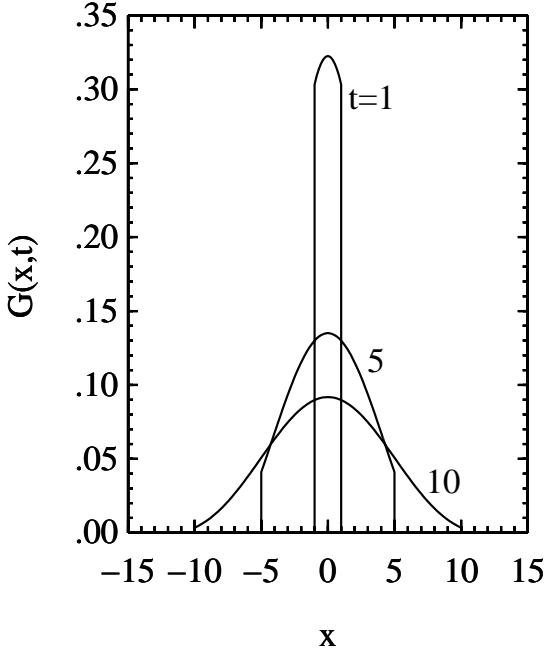}
\caption{Green function $G(x,t)$ plotted vs.\ $x$ at $t=1$, $5$, and $10$.
Note the discontinuities at $x=\pm t$.}
\label{fig-greeng}
\end{figure}

\begin{figure}
\includegraphics[width=70mm]{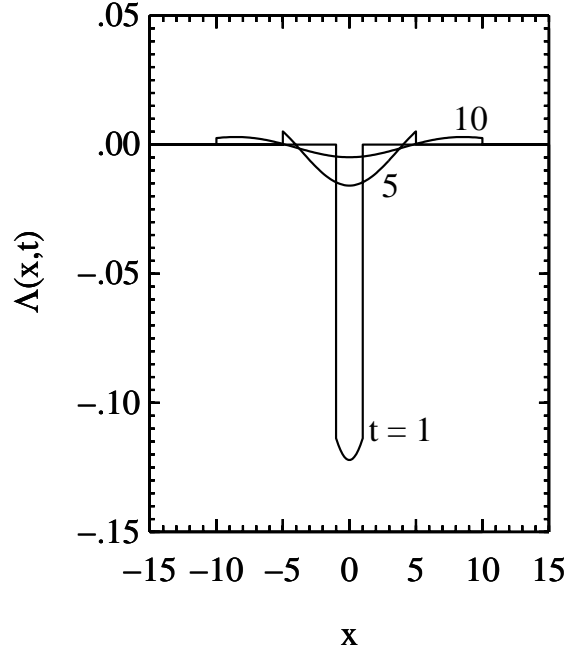}
\caption{Function $\Lambda(x,t)$ plotted vs.\ $x$ for $t=1$, $5$, and $10$.}
\label{fig-lambda}
\end{figure}

\begin{figure}
\includegraphics[width=70mm]{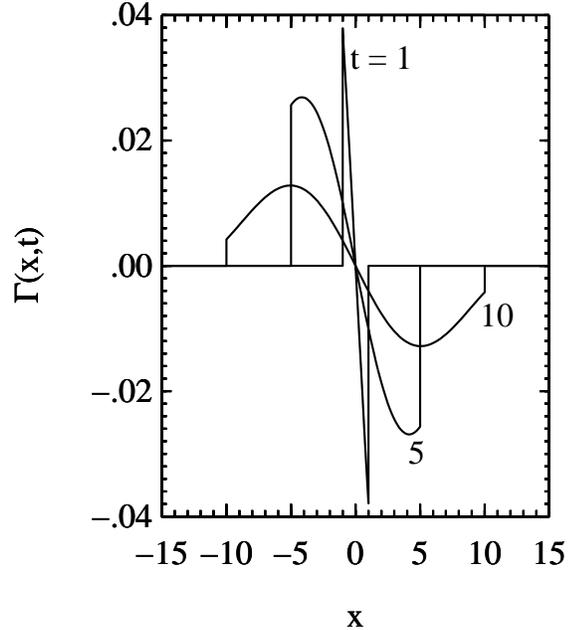}
\caption{Function $\Gamma(x,t)$ plotted vs.\ $x$ for $t=1$, $5$, and $10$.}
\label{fig-gamma}
\end{figure}

The Green function $G$ is calculated in Appendix~\ref{app-green}.
We find that
\be
G\xt = \half\,e^{-t/2}\,I_0\left(\xi/2\right)\,\Pi\left(\frac{x}{2t}\right),
\ee
where $I_0$ it the modified Bessel function of order zero,
\be
\xi \equiv \sqrt{t^2-x^2},
\ee
and the rectangle function,
\be
\Pi\left(x\right) \equiv \left\{
\begin{array}{rl}
1 & {\rm if}~\left|x\right|<1/2 \\
  &                             \\
0 & {\rm if}~\left|x\right|>1/2 \\
\end{array}
\right.,
\ee
insures causality.
In Fig.~\ref{fig-greeng} we plot $G$ vs.\ $x$ at selected times.

%
%
Differentiating $G$ yields the other Green functions,
\be
\Gdot\xt = \Lambda\xt +\half\,e^{-t/2}\delta(x+t) + \half\,e^{-t/2}\delta(x-t)
\label{eq-Gdotxt}
\ee
and
\be
\Gprm\xt = \Gamma\xt +\half\,e^{-t/2}\delta(x+t) - \half\,e^{-t/2}\delta(x-t),
\label{eq-Gprmxt}
\ee
where
\be
\Lambda\left(x,t\right) \equiv
-\frac{1}{2}G\xt
+\frac{1}{4}t\,e^{-t/2}\,
\left[\frac{I_1\left(\xi/2\right)}{\xi}\right]\,
\Pi\left(\frac{x}{2t}\right)
\label{eq-Lambda}
\ee
and
\be
\Gamma\left(x,t\right) \equiv
-\frac{1}{4}x\,e^{-t/2}\,
\left[\frac{I_1\left(\xi/2\right)}{\xi}\right]\,
\Pi\left(\frac{x}{2t}\right).
\ee
Plots of $\Lambda$ and $\Gamma$ are given in
Fig.~\ref{fig-lambda}--\ref{fig-gamma}.

It is useful to know the normalization of the Green
functions (e.g., to check the numerical evaluation of
convolution integrals).
It is easy to show that
\be
\int_{-\infty}^{+\infty}\ dx\ G\xt = 1-e^{-t},
\ee
\be
\int_{-\infty}^{+\infty}\ dx\ \dot{G}\xt = e^{-t},
\ee
and
\be
\int_{-\infty}^{+\infty}\ dx\ G^{\prime}\xt = 0.
\ee

\subsection{Large Time Behavior}

%
%
%
The discussion of the dispersion relations (\S\ref{sec-dispersion})
suggests that diffusion dominates the physics on large length scales.
Because the damping rate of the diffusion mode increases with
$k$ (Fig.~\ref{fig-dispi}),
this should be reflected in the large-time behavior of the Green functions.
In Appendix~\ref{app-asympt} we show that
\be
G\xt \longrightarrow G_{\rm diff}\xt \equiv
\frac{1}{\sqrt{4\pi t}}\,e^{-\frac{x^2}{4t}}
~~~~{\rm as}~t \longrightarrow\infty.
\label{eq-gasympt}
\ee
Consistent with the discussion in \S\ref{sec-dispersion},
$G\xt$ approaches the Green function for the diffusion equation
with unit diffusion coefficient.
We also show in App.~\ref{app-asympt} that the other Green functions
are smaller than $G$ at large times, with
\be
G:\Gprm:\Gdot ~\sim~ 1:t^{-1/2}:t^{-1}
\label{eq-gorders}
\ee
(see Fig.~\ref{fig-asympt}).

The asymptotic forms of the Green functions lead to a particularly
simple result for the solution when driving is absent.
If one replaces $G$ with $G_{\rm diff}$ and uses expression
(\ref{eq-gorders}) to omit all but the nonvanishing terms of largest
order in the homogeneous solution, the latter becomes
\be
\bh\xt \approx \left<G_{\rm diff}\left|\bz\right.\right>
\label{eq-bhasympt}
\ee
and
\be
\uh\xt \approx -\left<G_{\rm diff}^{\prime}\left|b_0\right.\right>.
\label{eq-uhasympt1}
\ee
Equation (\ref{eq-bhasympt}) indicates that the magnetic field
satisfies a diffusion equation when $t\gg 1$, as expected.
Equation (\ref{eq-uhasympt1}) is understood by noting that
\be
\left<G_{\rm diff}^{\prime}\left|b_0\right.\right> = 
\frac{\partial}{\partial x}\,\left<G_{\rm diff}\left|\bz\right.\right>,
\ee
so that eq.~(\ref{eq-uhasympt1}) is the same thing as
\be
\uh\xt \approx -b_{\rm h}^{\prime}\xt.
\label{eq-uhasympt}
\ee
The last expression says that the magnetic and drag forces
on the charged fluid balance one another
[cf.\ eq.~(\ref{eq-momentum-lin}) with
$\un=0$].
This is also expected: when $t\gg 1$, the inertia of the ions
becomes negligible and they undergo force-free motion.
\begin{figure}
\includegraphics[width=60mm]{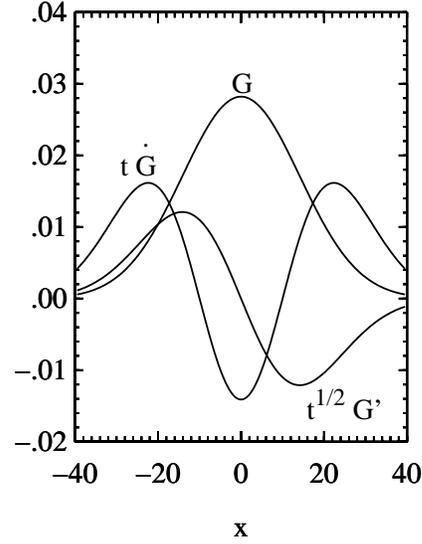}
\caption{The functions $G$, $t^{1/2}\Gprm$, and $t\Gdot$
plotted vs.\ $x$ for $t=100$.}
\label{fig-asympt}
\end{figure}

\subsection{Small Time Behavior}
\label{sec-nofriction}

When $t\ll 1$, the effects of friction are small and
the results of \S\ref{sec-drivenwaves} for driven waves should reduce
to the corresponding results for ideal MHD.
The latter can be found in a form suitable for comparison by setting
$u=\un=0$ in the linearized momentum equation and retracing the steps
leading to eq.~(\ref{eq-rhsol})--(\ref{eq-uhsol}) (for the homogeneous solution)
and eq.~(\ref{eq-rpsol})--(\ref{eq-upsol}) (particular solution).
The calculation is very straightforward
and we simply give the results:
\be
b_{\rm nf}\xt = 
        \left<\dot{G}_{\rm nf}\left|\bz\right.\right>
       -\left<G^{\prime}_{\rm nf}\left|\uz\right.\right>,        
\label{eq-bnf}
\ee

and
\be
u_{\rm nf}\xt = 
        \left<\dot{G}_{\rm nf}\left|\uz\right.\right>
       -\left<G^{\prime}_{\rm nf}\left|\bz\right.\right>,      
\label{eq-unf}
\ee
where the ``frictionless Green functions'' are
\be
\dot{G}_{\rm nf}\xt = \half\,\delta(x+t)+\half\,\delta(x-t)
\ee
and
\be
G^{\prime}_{\rm nf}\xt = \half\,\delta(x+t)-\half\,\delta(x-t).
\ee

We wish to compare expressions (\ref{eq-bnf}) and (\ref{eq-unf})
to the corresponding solution with friction in the limit $t\ll 1$.
The particular solution obviously goes to zero as $t\rightarrow 0$.
To find the limit of the homogeneous solution, we note
that any convolution integral approaches zero if its kernel
is $G$, $\Lambda$, or $\Gamma$, because the amplitude of each of these 
functions
remains finite as $t\rightarrow 0$, and is zero for 
$\left|x\right| > t$.
Retaining only the delta function kernels 
(with $\exp\left(-t/2\right)  \approx 1$ for $t\ll 1$) is equivalent to
making the replacements
\be
G\xt \rightarrow 0,
\ee
\be
\Gdot\xt \rightarrow \dot{G}_{\rm nf}\xt,
\ee
and
\be
\Gprm\xt \rightarrow G^{\prime}_{\rm nf}\xt.
\ee
in eq.~(\ref{eq-bhsol})--(\ref{eq-uhsol}).
This shows that the physics of driven waves reduces to ideal
MHD in the appropriate limit.

\begin{figure}
\includegraphics[width=84mm]{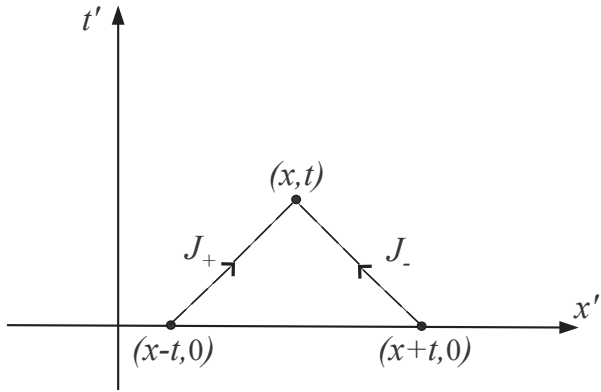}
\caption{In ideal MHD the Riemann invariants $J_{\pm}$ are conserved
quantities which are transported along characteristics. In the
presence of friction they are transported along the same
characteristics but not conserved.}
\label{fig-riemann}
\end{figure}

A comparison of the Green functions for waves with and without
friction also shows how ion-neutral friction alters the physics
of wave propagation.
This is best illustrated by examining the solutions in terms of the
Riemann invariants,
\be
J_-\xt \equiv b\xt-u\xt
\ee
and
\be
J_+\xt \equiv b\xt+u\xt.
\ee
In terms of $J_-$ and $J_+$ the solution for ideal MHD becomes
\be
J_-\xt = J_-\left(x+t,0\right)
\label{eq-Jmcons}
\ee
and
\be
J_+\xt = J_+\left(x-t,0\right),
\label{eq-Jpcons}
\ee
where we have used the fact that
\be
\left<f\left|\delta\left(x\pm t\right)\right.\right>
=
f\left(x\pm t\right)
\ee
to evaluate the convolutions.
The analogous solution for driven waves is
\be
J_-\xt = J_-\left(x+t,0\right)\,e^{-t/2}+
         \left<G\left|\bz\right.\right>+
         \left<\Lambda\left|J_-\right.\right>+
         \left<\Gamma\left|J_+\right.\right>
\label{eq-Jmfric}
\ee
and
\be
J_+\xt = J_+\left(x-t,0\right)\,e^{-t/2}+
         \left<G\left|\bz\right.\right>-
         \left<\Gamma\left|J_+\right.\right>+
         \left<\Lambda\left|J_-\right.\right>.
\label{eq-Jpfric}
\ee

Expressions (\ref{eq-Jmcons}) and (\ref{eq-Jpcons}) say
that the Riemann invariants are conserved quantities
in ideal MHD. The former is transported along the characteristic curve
$dx/dt=-1$ and the latter along $dx/dt=+1$ (Fig.~\ref{fig-riemann}).
It follows that the solution at each spacetime point \xt\ depends
only on the initial conditions at the points $(x-t,0)$ and $(x+t,0)$ 
(Fig.~\ref{fig-riemann}). This is the essence of wave propagation.
When friction is present, the situation is similar yet not identical.
The first term on the RHS of eq.~(\ref{eq-Jmfric}) and 
eq.~(\ref{eq-Jpfric}) may be interpreted to mean that the Riemann
``invariants'' are still transported along characteristics, but
with exponential attenuation. That is, they represent wave propagation 
with damping. However the presence of additional terms shows that this 
is not the whole story.
For example, the nonzero widths of the convolution kernels on
each RHS imply that the solution at \xt\ depends on the
initial conditions over the entire spatial interval $\left(x-t,x+t\right)$.
This is the signature of diffusion.

The physics of wave propagation, diffusion, and the transition
from wavelike to diffusive behavior is apparent
in the plots of the Green functions (Fig.~\ref{fig-greeng}--\ref{fig-gamma}).
Wave fronts are located by the jumps at $x=\pm t$, propagating with 
velocities of $\pm 1$.
Diffusion is implied by the finite widths of the Green functions.
The transition from wave propagation to diffusion occurs
at times of order unity, when the jumps have decayed conspicuously.
At much greater times the jumps become invisible and
all vestiges of wave propagation disappear.

\subsection{Steady Driving}
\label{sec-steady}

The physics of driven waves simplifies dramatically
when the frictional driving is steady.
Since we are interested in
time scales where the neutral flow is approximately steady,
it is worth discussing this case in some detail.
Only the particular solution needs to be considered;
the homogeneous solution depends only on the initial conditions,
and so requires no modification.

If the neutral flow is steady then it is possible to write
\be
\un\xt = D\left(x-\ug t\right),
\ee
where $D(x)$ is the neutral velocity profile
and \ug\ is the pattern velocity.\footnote{\label{ftn-GI}Note that the linearized
equations of motion are not invariant under Galilean transformations:
they are valid only in the frame where the undisturbed gas far
upstream is at rest. The pattern velocity is evaluated in this frame.}
Then the Fourier 
transform of \un\ has the form
\be
\unhat\kt = \Dhat(k)\,\emikugt
\label{eq-unhatsteady}
\ee
and the simple time dependence of \unhat\ makes it possible
to find the expansion coefficients $C_m$ explicitly.
Substituting expression (\ref{eq-unhatsteady}) into equations
(\ref{eq-Cmsolution}) and (\ref{eq-Cpsolution}) and evaluating
the integrals gives
\be
C_-\,\emiwmt = \Dhat(k)\,\Shatmn(k)\,\left[\emikugt - \emiwmt\right],
\label{eq-Cmnsteady}
\ee
and
\be
C_+\,\emiwpt = \Dhat(k)\,\Shatpl(k)\,\left[\emikugt - \emiwpt\right],
\label{eq-Cplsteady}
\ee
where
\be
\hat{S}_{\pm}(k) \equiv
\pm\frac{\hat{Q}}{R}+\Phat,
\label{eq-spmdef}
\ee
%
%
%
\be
\Qhat(k) \equiv 
-\frac{\imath}{4} 
\left[
\frac{
2k-\imath u_{\rm g}
}{
k\left(1+u_{\rm g}^2\right) -\imath u_{\rm g}}
\right]
\approx
-\frac{\imath}{4} 
\left(
\frac{
2k-\imath u_{\rm g}
}{
k-\imath u_{\rm g}}
\right),
\label{eq-qhatdef}
\ee
\be
\Phat(k) \equiv 
\frac{1}{2} \left[
\frac{-\imath u_{\rm g}
}{
k\left(1+u_{\rm g}^2\right) -\imath u_{\rm g}}
\right]
\approx
\frac{1}{2} \left(
\frac{-\imath u_{\rm g}
}{
k -\imath u_{\rm g}}
\right).
\label{eq-phatdef}
\ee
}
Having found $C_-$ and $C_+$, we only need to calculate the
inverse transform of expression~(\ref{eq-yhatpdef}) to find the
particular solution.
Evidently the latter contains two parts:
terms in $C_{\pm}$ that are proportional to
$\exp\left(-\imath ku_{\rm g}t\right)$ represent
steady flow of the {\it charged}\/ fluid,
which must eventually result from steady driving by the neutrals.
The other terms in $C_{\pm}$ represent the transient response to driving.

First we evaluate the velocity, \up.
Substituting expressions (\ref{eq-Cmnsteady}) and (\ref{eq-Cplsteady})
into eq.~(\ref{eq-yhatpdef}) gives
\be
\uhatp = \uhatps + \uhatpt,
\ee
where
\be
\uhatps \equiv \Dhat\,\left(\Shatmn+\Shatpl\right)\,\emikugt
\label{eq-uhatps}
\ee
and
\be
\uhatpt \equiv -\Dhat\,
\left(\Shatmn\,\emiwmt+\Shatpl\,\emiwpt\right)
\label{eq-uhatpt}
\ee
are the steady and transient parts, respectively.
Now the inverse transform of expression (\ref{eq-uhatps}) is
\be
\ups\xt = \int_{-\infty}^{+\infty}\ dk\,
\Shat(k)\,\Dhat(k)\,
e^{\imath k \left(x-u_{\rm g}t\right)},
\label{eq-upsint}
\ee
where
\be
\Shat(k) \equiv \Shatmn+\Shatpl = \frac{-\imath\ug}{k-\imath\ug}.
\ee
The inverse transform of \Shat\ is evaluated by simple
contour integration to find
\be
S(x) = \left\{
\begin{array}{ll}
0                                & {\rm if}~\ug x<0 \\
                                 &              \\
2\pi\left|\ug\right|\,\exp\left(-\ug x\right) & {\rm if}~\ug x>0 \\
\end{array}
\right. .
\ee
If we use this result and the Convolution Theorem to evaluate
expression (\ref{eq-upsint}), we obtain the steady part of
the ion velocity:
\be
\ups\xt = \Dtilde\left(x-\ug t\right),
\label{eq-upsteady}
\ee
where
\be
\Dtilde(x) \equiv \frac{1}{2\pi}\,\int_{-\infty}^{+\infty}\ dx^{\prime}\,
S\left(x^{\prime}\right)\,
D\left(x-x^{\prime}\right).
\label{eq-Dtilde}
\ee
This is a remarkably simple result.
If the neutral flow is steady, then the charged fluid will eventually
undergo steady flow with the same pattern velocity, consistent with common sense.
When both flows are steady, the velocity of the charged fluid is just the
convolution of the driving force, $D$, with an exponential response
function, $S$.

Now consider the transient part of the ion velocity.
Substituting expressions (\ref{eq-spmdef})--(\ref{eq-phatdef})
into equation (\ref{eq-uhatpt}), we find after routine algebra
that
\be
\uhatpt\kt = 2\pi \left[
\left(2\imath\Qhat-\Phat\right)\Dhat\Ghat
-2\Phat\Dhat\Gdothat
\right].
\ee
Inverting the transform is performed by a twofold application of
the Convolution Theorem once the inverse transforms of $\Qhat$
and $\Phat$ are known.
We find that
\be
P(x) = \frac{1}{2} S(x)
\ee
and
\be
Q(x) = -\frac{\imath}{2u_{\rm g}}\,S^{\prime}(x)-\frac{\imath}{4} S(x).
\ee
Using these results and the Convolution Theorem yields
\be
\upt\xt = 
\left<G\left|D\right.\right>
-
\left<G\left|\Dtilde\right.\right>
-
\left<\Gdot\left|\Dtilde\right.\right>.
\label{eq-uptransient}
\ee

The transform of the magnetic field perturbation is
\be
\bhatp = 
\frac{k}{\omega_-}\,\Cmn\,\emiwmt
~+~
\frac{k}{\omega_+}\,\Cpl\,\emiwpt.
\ee
The inverse transforms are obtained by steps very similar to those
used to find the velocity.
We omit the details and simply state the result:
\begin{eqnarray}
\label{eq-steady-rbsoln}
\bp\xt = &  \Dtilde\left(x-u_{\rm g}t\right)/\ug
                    -\left<G\left|\Dtilde/\ug\right.\right> \nonumber \\
                  & -\left<\Gdot\left|\Dtilde/\ug\right.\right>
                    +\left<G\left|D\ug\right.\right>.\\
\nonumber
\end{eqnarray}

\section{A Cloud-Cloud Collision}
\label{sec-collision}

As an application of the methods developed in \S\ref{sec-drivenwaves}, 
we consider the collision of two identical, uniform, semi-infinite 
clouds. The relative motion is taken to be along \vecxhat\
and the magnetic fields to be everywhere along \veczhat.
Prior to the collision the free surface of each cloud 
is a plane normal to \vecxhat.
The collision occurs at $t=0$, when the free surfaces touch
at the contact discontinuity, $x=0$.
We take the relative velocity of the clouds to be $\Delta v=20$\,\kms\
and adopt initial conditions appropriate for a dense core:
the neutral fluid in each cloud is pure molecular hydrogen
with $\enn=2\times 10^4$\,\cmMMM,
the charged fluid has $\emi=25$\,amu, $\eni/\enn = 3\times 10^{-8}$,
and the unperturbed magnetic field is $B_0=50\,\mu$G.
Then the characteristic speeds, time- and length
scales are $\viAz=894$\,\kms,
$\tauin=1.1 \times 10^{-2}$\,yr, $\tauni=3.0 \times 10^4$\,yr,
and $\viAz\,\tauin = 3.2 \times 10^{13}$\,cm.

The homogeneous solution depends only on the initial conditions,
which are
\be
r_0(x) = b_0(x) = 0
\ee
for the density and field perturbations.
We adopt a reference frame (the ``CM frame'') where
the clouds approach with equal and opposite velocities. 
In this frame the initial velocity perturbation is
%
%
\be
u_0(x) = -\frac{1}{2}\Delta u\,{\rm sgn}(x),
\label{eq-precursoric}
\ee
where
\be
\Delta u = \frac{\Delta v}{v_{\rm iA0}} \approx .022
\ee
and
\be
{\rm sgn}(x) \equiv \left\{
\begin{array}{cl}
-1 & x<0 \\
   &     \\
+1 & x>0 \\
\end{array}
\right. .
\ee
Using these initial conditions, we
calculated the homogeneous solution
by evaluating expressions (\ref{eq-bhsol}) and 
(\ref{eq-uhsol}).\footnote{Taking care to evaluate
expressions  (\ref{eq-bhsol}) and (\ref{eq-uhsol}) in
reference frames where they are valid. See footnote \ref{ftn-GI}.}

\begin{figure}
\includegraphics[width=70mm]{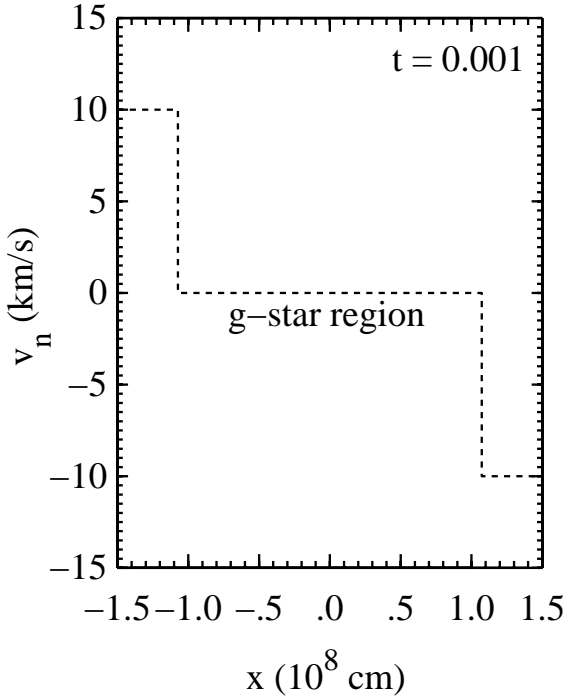}
\caption{Velocity of the neutral fluid in the CM frame at the extremely
early time $t=0.001$.
The discontinuities are forward ($x>0$)
and reverse ($x<0$) J shocks. They propagate away from the contact
discontinuity, $x=0$, at a constant speed of $3.3$\,\kms.}
\label{fig-nflow}
\end{figure}
\begin{figure}
\includegraphics[width=70mm]{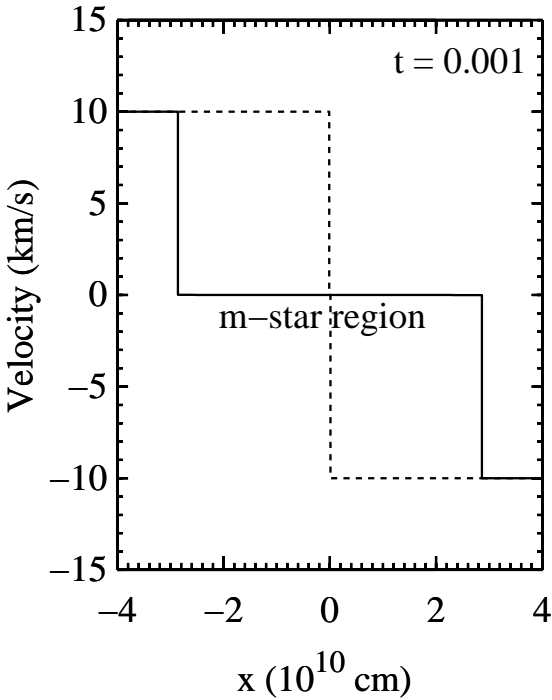}
\caption{Velocity profiles of the charged (solid) and 
neutral (dashed) fluids at the extremely early time $t=0.001$.
The g-star region is only
$\approx 10^8\,{\rm cm}$ wide at this time and so unresolved.}
\label{fig-uhyper}
\end{figure}
\begin{figure}
\includegraphics[width=70mm]{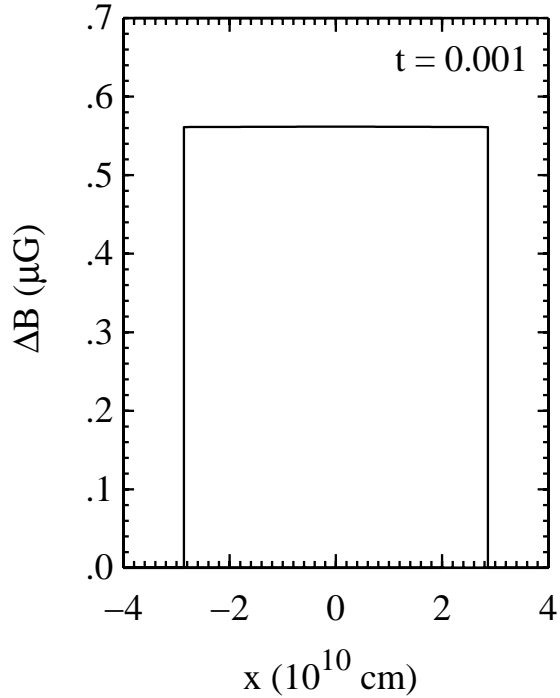}
\caption{Magnetic field perturbation $\Delta B \equiv B-B_0$
at the extremely early time $t=0.001$.}
\label{fig-bhyper}
\end{figure}

The particular solution depends on the velocity profile of the neutral 
fluid, which is determined 
by solving Euler's equations. This is 
straightforward as the example considered here is just a special case 
of the Riemann problem in gas dynamics. Using the Riemann solver of 
\cite{Toro99}, we calculated $\vn\xt$ for an adiabatic ideal gas with 
$\gamma=5/3$. We set the temperature of the unperturbed neutral gas 
somewhat arbitrarily to $T_{\rm n,0}=10$\,K;
%
%
that is, we did not find $T_{\rm n,0}$ self-consistently
by requiring thermal balance in the unperturbed state. (Since the shocks in this
example are very strong, the value of $T_{\rm n,0}$ hardly affects the solution.)
The neutral flow is steady.
It consists of forward- and reverse J shocks propagating with
constant velocities in the $\pm x$ directions. The shock velocities are
$\pm 3.3$\,\kms\ in the CM frame, corresponding
to identical shock speeds of 13.3\,\kms\ relative to the upstream gas.
Having found \vn\xt, we obtained the particular solution
from expressions 
(\ref{eq-upsteady}), (\ref{eq-uptransient}), and (\ref{eq-steady-rbsoln}).

The velocity of the neutral fluid is plotted in Fig.~\ref{fig-nflow}
at the extremely early time $t=0.001$  ($\leftrightarrow 300$\,s!).
The region between the J fronts
(the ``g-star region'') contains
neutral gas which has been swept up, heated, and compressed by the shocks.
Because these shocks are very strong (with Mach numbers $>50$), the swept-up
gas is denser than the undisturbed gas by a large factor ($\approx 4$),
in gross violation of our assumption that \enn\ is constant.
Moreover the g-star region is so hot ($\Tn \approx 8,000$\,K)
that radiative cooling is important even on time scales $\ltsim 1$\,yr.
In a separate paper we describe numerical
calculations that allow for variations in \enn, radiative
cooling, and other (e.g., nonlinear) effects.
Here we temporarily forego these complications and simply warn the reader
that our example may not be correct in detail.

The velocity profile of the charged fluid at $t=0.001$ is plotted in
Fig.~\ref{fig-uhyper}.
At this time the effects of friction are negligible,
so the flow in Fig.~\ref{fig-uhyper}
is just the magnetohydrodynamic analog of the flow
in Fig.~\ref{fig-nflow}.
The MHD flow has a growing ``m-star region''
bounded on either side by a moving J front.
In both the gas dynamic and MHD flows, a particle of fluid passing
through a J front is brought to rest by an impulsive force inside the front;
the force is collisional in the first case and magnetic in the second.
In the gas dynamic flow, the collisional force vanishes outside
the J fronts so the neutral fluid remains at rest inside
the g-star region (Fig.~\ref{fig-nflow}).
In the MHD flow, the magnetic force vanishes outside the J
fronts but friction does not.
The charged fluid appears to be at rest in the m-star region
(Fig.~\ref{fig-uhyper}) only because $t\ll \tauin$.

The magnetic field at $t=0.001$ is plotted in Fig.~\ref{fig-bhyper}.
The m-star region is conspicuous as a ``magnetic core'' of
compressed fluid and magnetic field centered on the contact discontinuity.
The compression is approximately uniform because the velocity
gradient in the m-star region is approximately zero.
The compression ratio is small ($\approx 1.01$) because
the ``shocks'' in the charged fluid have ion \Alfven\ Mach
numbers of only $\approx 0.01$.
Consistent with the linear analysis,
the velocity jumps ($\delta v_i = \pm 10$\,\kms)
and magnetic field jumps ($\delta B = 0.562$\,\muG) obey the relation
\be
\frac{\left|\delta B\right|}{B_0} = 
\frac{\left|\delta v\right|}{v_{\rm iA0}}
\ee
for linear magnetosound waves.
We note also that the mathematical approach we have adopted
treats discontinuities exactly (i.e., without
smoothing) and that discontinuties propagate at the
correct speeds.

\begin{figure}
\includegraphics[width=70mm]{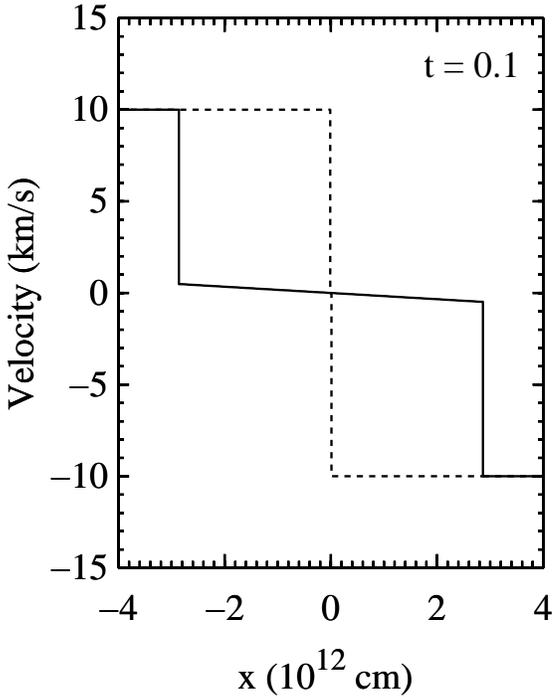}
\caption{Velocity profiles of the charged (solid) and 
neutral (dashed) fluids at $t=0.1$.
The horizontal scale is too coarse to resolve the g-star region.}
\label{fig-uearly}
\end{figure}

\begin{figure}
\includegraphics[width=70mm]{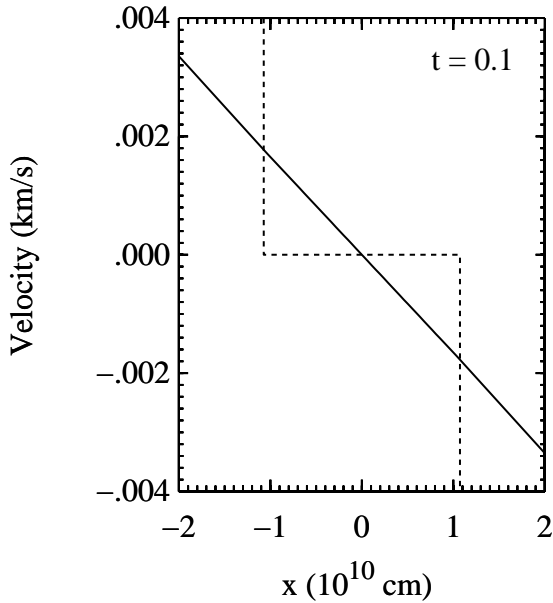}
\caption{As in Fig.~\ref{fig-uearly} but plotted on a scale
that resolves the g-star region.}
\label{fig-uearlyf}
\end{figure}

\begin{figure}
\includegraphics[width=70mm]{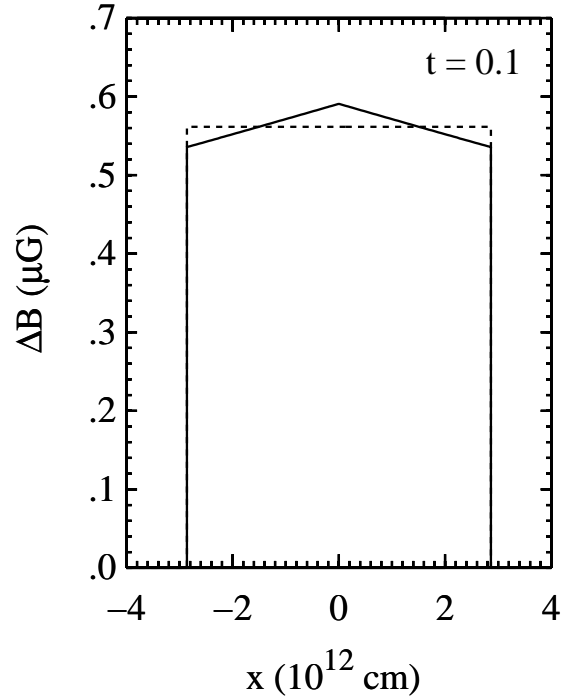}
\caption{Solid: Magnetic field perturbation at $t=0.1$.
The horizontal scale is too coarse to resolve the g-star region.
Dashed: Result of an identical calculation with friction turned off.
Notice that the area under each curve is the same.}
\label{fig-bearly}
\end{figure}

Figures \ref{fig-uearly}--\ref{fig-bearly} describe the flow
at $t=0.1$ ($\leftrightarrow 3\times 10^4$\,s),
when the J fronts are at $x=\pm 1.07 \times 10^{10}$\,cm
in the neutral fluid and  $x=\pm 2.86\times 10^{12}$\,cm in the charged fluid.
Departures from ideal MHD are now visible as slight reductions in
$\delta\vi$ (to $\approx \pm 9.5$\,\kms )
and $\delta B$ (to $\approx 0.536$\,\muG ).
The ions inside the m-star region are moving. They flow toward the 
contact discontinuity with $\vi \approx \pm 0.5$\,\kms\ just
downstream from the J fronts, decreasing to $\vi=0$ at $x=0$
(Fig.~\ref{fig-uearlyf}).
The compression inside the m-star region is no longer uniform;
however the density/magnetic field gradient is almost constant, with $B$
and $\rho$ increasing toward $x=0$ (Fig.~\ref{fig-bearly}).

The qualitative properties of the flow at $t=0.1$ follow from simple physics.
To see this, it is useful to note that a particle of charged fluid moving
with $\vi=0.5$\,\kms\  would travel only $\sim 10^9$\,cm in
$3\times 10^4$\,s.
Each point on the solid curves in Fig.~\ref{fig-uearly}--\ref{fig-bearly}
therefore labels a fluid particle which has not moved appreciably
since $t=0$.
Now consider the histories of various fluid particles,
starting from the initial state $t=0.001$ when the effects
of friction on the flow were negligible.
More specifically, consider the forces  at $t=0.001$ on particles that
are inside the m-star region excluding the J fronts.
All of these particles have $\vi=0$ at $t=0.001$ (Fig.~\ref{fig-uhyper})
and the magnetic force on each vanishes (Fig.~\ref{fig-bhyper}).
The friction force is small inside the g-star region ($\vn-\vi \approx 0$)
but not outside ($\vn-\vi \approx \pm 10$\,\kms ).
Since the net force is small inside the g-star region, we expect
the charged fluid there to be almost at rest at later times.
This is true at $t=0.1$ (Fig.~\ref{fig-uearlyf}).
The net (=frictional) force is much larger outside the g-star region;
the sign of $\vn-\vi$ is such that the particles tend to speed up
and flow toward the contact discontinuity.
This explains the nonzero values and sign of \vi\ at $t=0.1$
(Fig.~\ref{fig-uearly}).

\begin{figure}
\includegraphics[width=70mm]{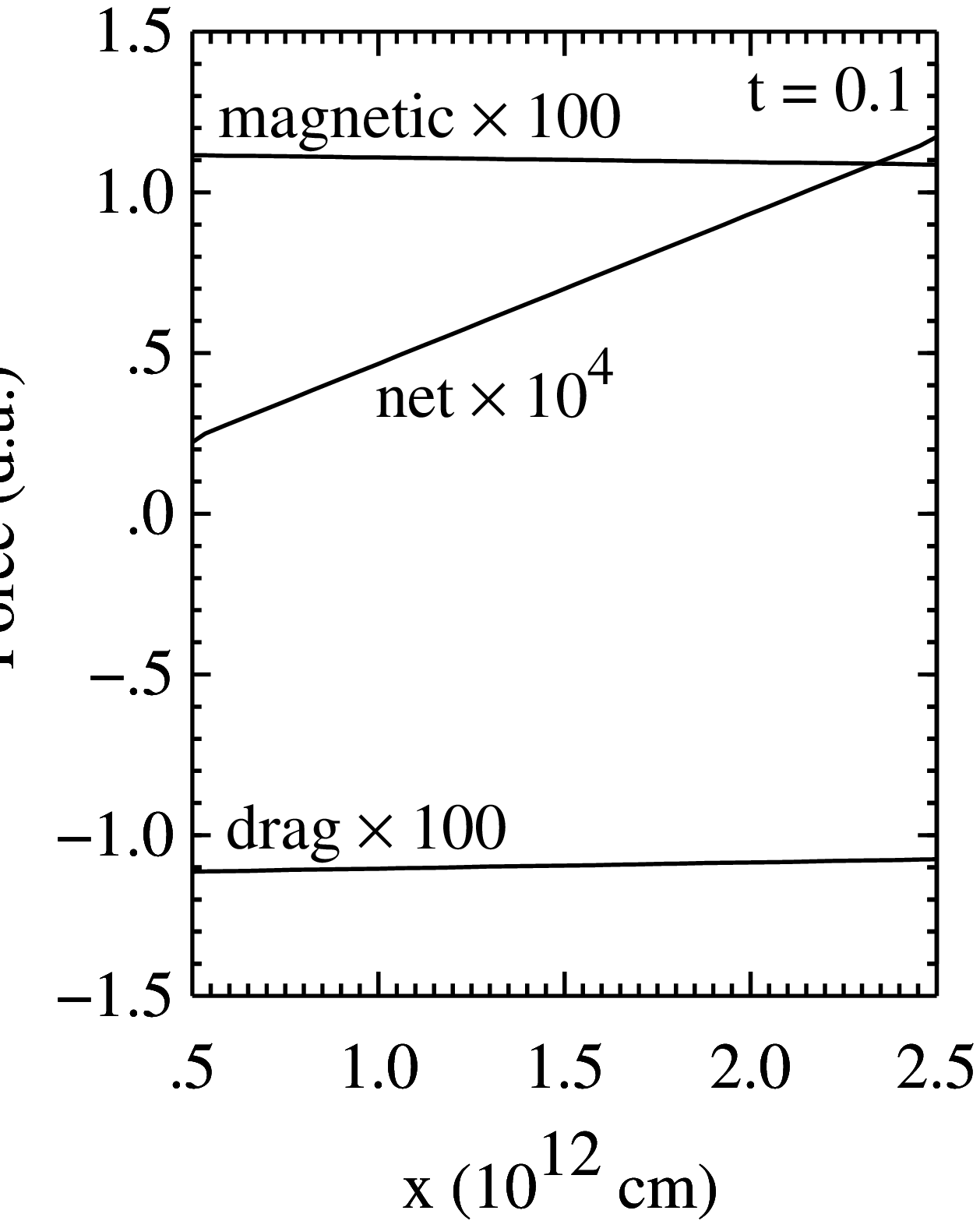}
\caption{Magnetic, drag, and net forces in dimensionless units at t=0.1.
Only points between the forward shocks in the neutral and charged
fluids are shown.}
\label{fig-fearly}
\includegraphics[width=70mm]{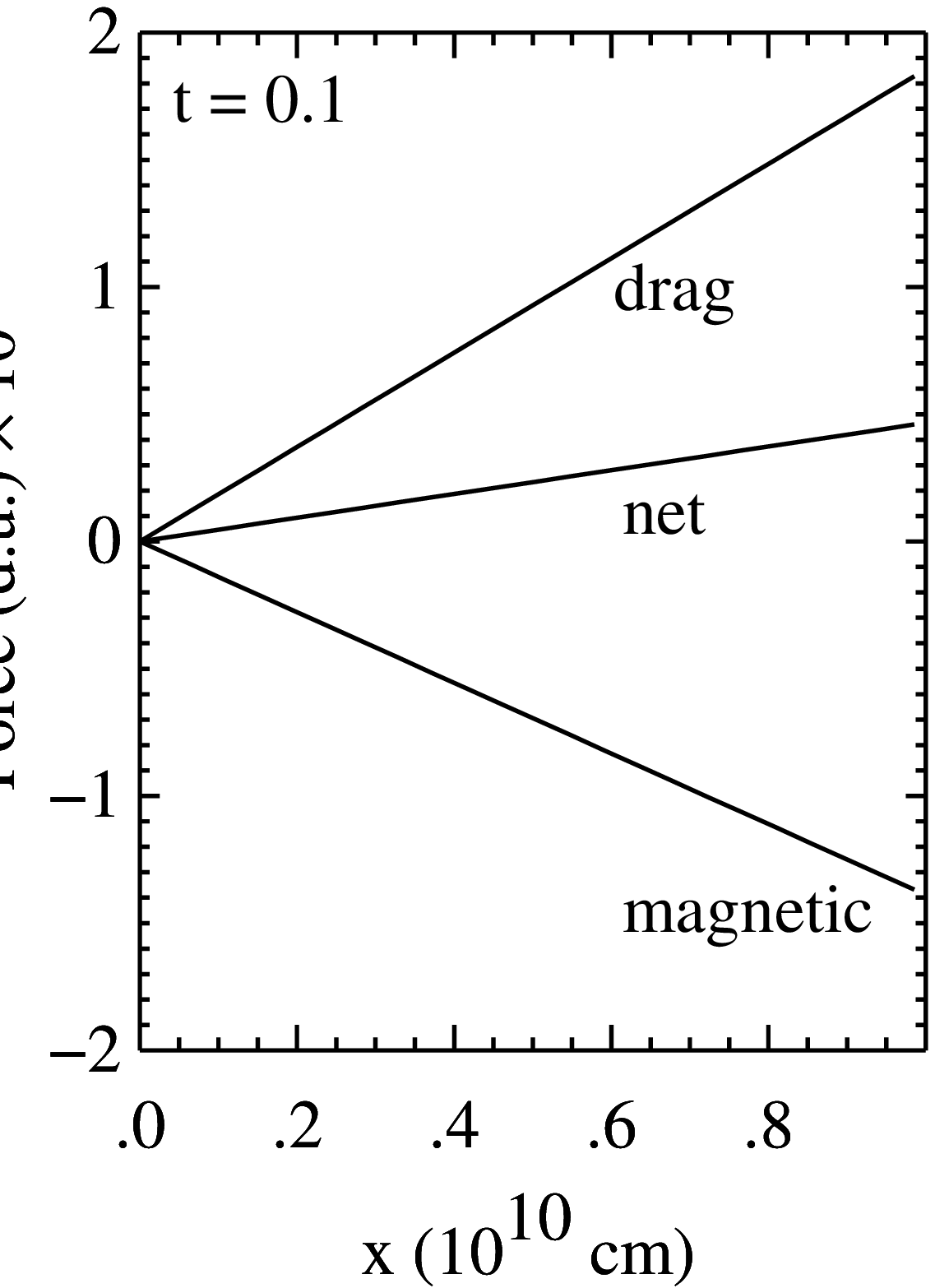}
\caption{As in Fig.~\ref{fig-fearly} but on a scale that resolves
the g-star region. Note force scale.}
\label{fig-fearlyf}
\end{figure}

However the preceding argument does not explain the velocity gradient in
Fig.~\ref{fig-uearly}, which would have the opposite sign if friction
was the only force at work.
This is because
particles close to $x=0$ have been 
accelerating longer than particles that have just emerged from the J 
front.
The argument is incomplete because, contrary to intuition, the
magnetic force is comparable to friction at $t=0.1$.
Indeed, the two forces 
differ by less than $\approx 1$\% (Fig.~\ref{fig-fearly}).
The magnetic force is caused by ambipolar diffusion.
The acceleration produced initially by friction produces
ion motions that transport magnetic field lines toward the contact
discontinuity.
Now the total magnetic flux threading the m-star region
is the same, at any given time, whether friction is present or
not (Fig.~\ref{fig-bearly}).\footnote{The total flux is just
the flux swept up by the J fronts, which depends only on $B_0$ and
the front velocities.
The latter are $\pm \viAz$ whether or not
friction is present.}
The motion of field lines toward $x=0$ must therefore
leave a deficit of field lines farther out
(relative to the frictionless case; compare the solid and dashed curves
in Fig.~\ref{fig-bearly}).
The result is a magnetic field gradient, the sign of which
has the magnetic force opposing friction everywhere.

The sign of the velocity gradient in Fig.~\ref{fig-uearly} can
be explained as follows.
The magnetic impulse delivered to a fluid particle inside a J front
is proportional to $\delta B$.
The transport of field lines away from the fronts reduces $\delta B$
and hence $\delta\vi$.
The result is a velocity gradient with the observed sign.
In fact there is a feedback mechanism at work: 
the reduction of $\delta\vi$ tends to make the velocity gradient {\it steeper},
which increases the rate of field line migration
(cf.\ eq.~\ref{eq-induction-lin}).
And so on. This explains why jumps in the fluid variables decay
{\it exponentially}\/ rather than (say) linearly with time.

As noted above, at $t=0.1$ magnetic and collisional forces 
are approximately equal between the advancing front 
and the contact discontinuity, differing by $\ltsim 1\%$. 
This may seem surprising, since one might think that balance
between forces could not occur for $t < 1$.
This is because $\tau_{\rm in}$ is basically the time it takes for an
ion to ``lose memory"
of its initial state through collisions with the neutrals.
It is readily shown that, in the absence of 
magnetic forces, the velocity of an ion in a uniform one-dimensional 
flow of neutrals is given by 
$v_{\rm i}(t) = v_{\rm i0}e^{-t/\tau_{\rm in}} 
+ v_{\rm n}\left(1 - e^{-t/\tau_{\rm in}}\right)$, where
$v_{\rm i0}$ is the ion's initial velocity.
Approximate balance between magnetic and drag forces, however, requires 
only that the two forces be comparable in magnitude, yielding a 
residual that is much smaller than the magnitude of the individual 
forces themselves. Effective force-balance in the ions behind the 
J front is then possible when the ratio of the net 
(linearized) acceleration relative to the drag acceleration becomes
negligible, i.e., when
$(\partial v_{\rm i}/\partial t)[\tau_{\rm in}/(v_{\rm n} - v_{\rm i})]
\ll 1$ 
(see eq. [2]). Taking $\partial v_{\rm i,ch}/\partial t \sim
v_{\rm i, ch}/t$, where $v_{\rm i,ch}$ is a characteristic
post-jump ion velocity, and using 
$v_{\rm n} - v_{\rm i}
\simeq v_{\rm n}$ in this region (see Fig. 12), this condition is 
equivalent to $t/\tau_{\rm in} \gg v_{\rm i,ch}/v_{\rm n}$. If the 
magnitude of the ion velocity in the post-shock region is sufficiently 
reduced by the magnetic impulse at the front so that 
$v_{\rm i,ch}/v_{\rm n} \ll 1$, it follows that near force-balance
in the ions can happen even for $t/\tau_{\rm in} < 1$. This is
exactly the situation that is depicted in Figs. 12 and 14: the
magnetic ``kick" the ions receive at the jump dramatically
reduces the ion velocity, resulting in 
$v_{\rm i,ch}/v_{\rm n} \ltsim 0.05$ behind the front, allowing 
near-equality of 
forces in that region at $t = 0.1$
($> 0.05$). For this case, the collisional drag of the streaming 
neutrals on the ions is almost enough to balance the magnetic force due
to the field gradient behind the front.

The ion motion is almost but not exactly force-free at $t=0.1$.
Outside the g-star region, gas drag pulls the ions toward the 
contact discontinuity and the magnetic force pushes them away (Fig.~\ref{fig-fearly}).
Because the magnetic force is slightly larger in magnitude the ions slow down,
to speeds which are almost but not quite zero when they pass into 
the g-star region (Fig.~\ref{fig-uearlyf}).
There both forces change sign (Fig.~\ref{fig-fearlyf});
the 
very small net force still tends to slow the ions down,
allowing them to come to rest at the contact discontinuity.

\begin{figure}
\includegraphics[width=70mm]{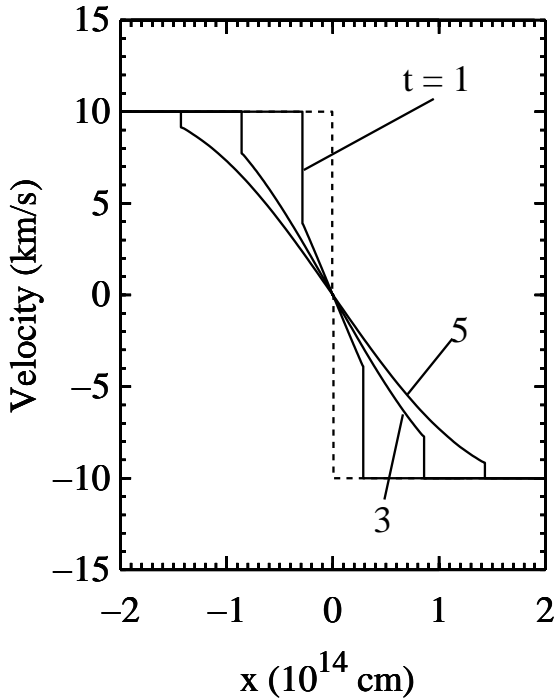}
\caption{Solid: Ion velocity at $t=1$, $3$, and $5$.
Dashed: Neutral velocity at t=5.}
\label{fig-umod}
\end{figure}

\begin{figure}
\includegraphics[width=70mm]{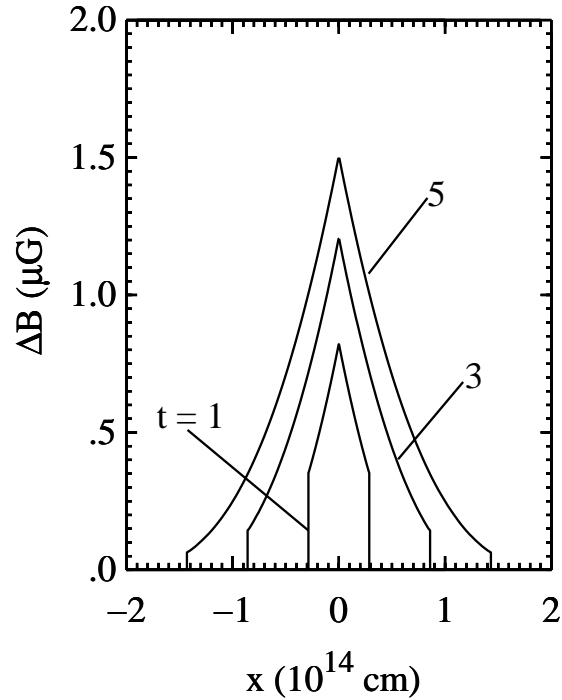}
\caption{Magnetic field perturbation at $t=1$, $3$, and $5$.}
\label{fig-bmod}
\end{figure}
Figures~\ref{fig-umod} and \ref{fig-bmod} describe
the flow at three times of order unity, the largest
of which is $\approx 0.05$\,yr.
The transport of field lines away from the J fronts
has attenuated the latter significantly, with a corresponding increase
in $B$ 
within the magnetic core.
These figures show the earliest phase in the formation of two
multifluid, MHD shock waves by the cloud-cloud collision.
At $t\sim 1$, each multifluid shock is comprised
of a J shock in the neutral fluid with a nascent magnetic precursor
extending upstream from its J front. Although we made some unrealistic 
assumptions in order to work the problem analytically
(no radiative cooling, \enn\ independent of $x$),
the mismatch between the time here ($<0.1$\,yr) and the
time to accelerate the neutral fluid ($\sim 10^4$\,yr)
leaves no doubt that the solution is qualitatively correct.

One does not expect the entire (charged+neutral) flow 
will approach a steady configuration until {\it much}\/ greater times
$\sim 10^4$\,yr.
However it is reasonable to suppose, given the large ``signal
speed'' in the charged fluid, that the magnetic precursors would
reach a quasi-steady state at shorter times.
This is clearly not the case at $t\sim 1$ where, for example,
the precursors are increasing in width (Fig.~\ref{fig-umod})
and the magnetic flux inside the core is growing (Fig.~\ref{fig-bmod}).
This evolving structure is determined mostly by waves 
propagating into the charged fluid.
Thus the linear scale ($\propto t$) is set by the steady speed of the J fronts.
In mathematical terms: the solution is ``mostly'' the homogeneous solution, 
i.e., the transient response of the charged fluid to the 
discontinuous initial conditions.
We refer to this evolutionary phase  henceforth as
the ``ion-electron transient.''
The ion-electron transient is unobservable; however it obviously
affects subsequent, possibly observable, phases.

\begin{figure}
\includegraphics[width=70mm]{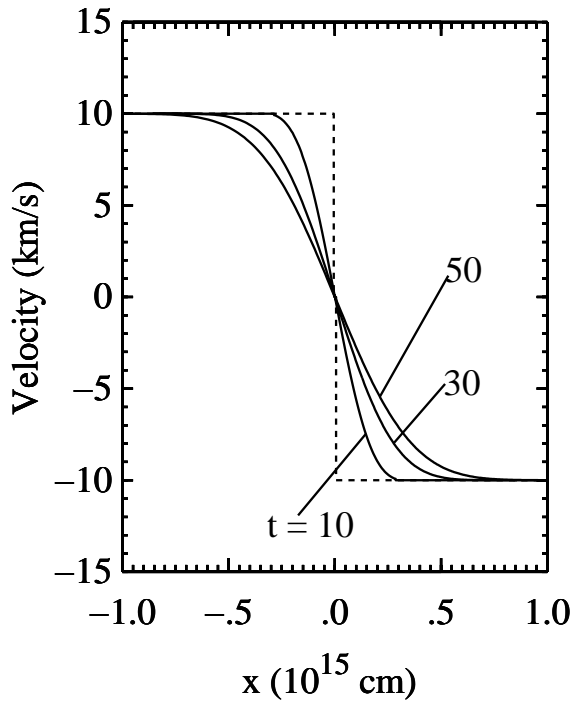}
\caption{Solid: Ion velocity at $t=10$, $30$, and $50$.
Dashed: Neutral velocity at t=50.}
\label{fig-ubig}
\end{figure}

\begin{figure}
\includegraphics[width=67mm]{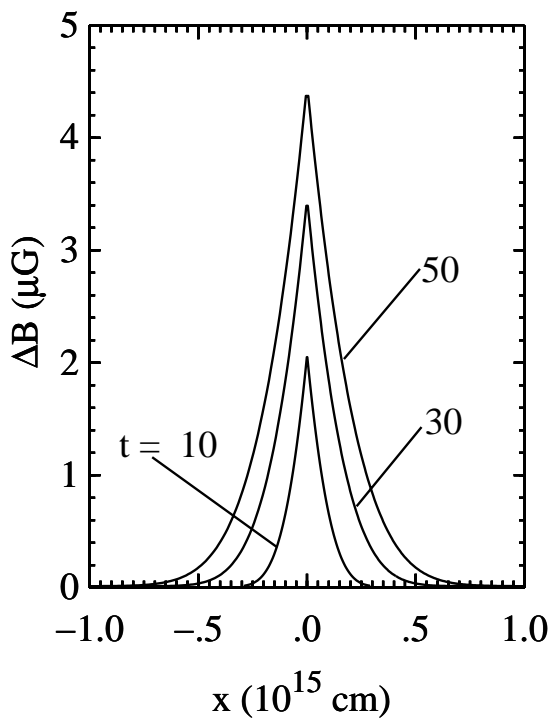}
\caption{Magnetic field perturbation at $t=10$, $30$, and $50$.}
\label{fig-bbig}
\end{figure}

Figures~\ref{fig-ubig} and \ref{fig-bbig} display
the flow at three times $\sim 10$, the largest of which
corresponds to $\approx 0.5$\,yr.
Now the solution is entirely the particular (=driven) solution.
The embedded neutral shocks influence the flow
mainly through the neutral velocity profile, which dictates
the spatial dependence of the friction force.
The fact that the neutral shocks are propagating, for example,
is relatively unimportant.
All traces of wave physics have disappeared but
the precursors are still evolving.
The flow in this phase is determined by the competition between
the advection of field lines inward, driven by the ion motions, and
the diffusion of field lines outward, due to the resulting
magnetic field gradient.
The dominance of diffusion is signalled by the linear scale,
which now is $\propto \sqrt{t}$.
We refer to this period as the ``diffusion phase'' of
multifluid shock formation.

It would be useful to know what physical effects terminate the
diffusion phase, whether this phase lasts long enough to be observable,
and whether the charged flow approaches a quasi-steady configuration.
Unfortunately, the solutions for $t > 50$ cannot be discussed
here because nonlinear effects start to become important
(e.g., Fig.~\ref{fig-bbig}).
Calculations in progress will use numerical methods to
include nonlinearity, radiative cooling, and variations
in the density of the neutral gas.
We are also exploring similarity solutions (for the diffusion phase)
of the nonlinear equations of motion.
If a similarity solution existed, it would rule out quasi-steady flow
of the charged particles.
It might also reveal which physical effects usher in the next
evolutionary phase.
The flow in Fig.~\ref{fig-ubig} and \ref{fig-bbig}
certainly appears to be self-similar.
It also satisfies the prerequisites:
there are no boundary conditions and the system has ``forgotten'' its
initial conditions.

\section{Summary}
\label{sec-summary}

This paper can be summarized as follows:

(i) Our objective was to understand basic physics governing
the formation of multifluid, MHD shock waves from plausible initial conditions.
We focused on the earliest stages of this process, which have not been
explored elsewhere.

(ii) We treated the plasma as separate fluids of charged and neutral
particles which are coupled by ion-neutral friction.
We exploited the large inertial mismatch between the neutral and
charged fluids to simplify the calculation.
On time scales $\ltsim 10^4$\,yr (typically), 
the neutral fluid evolves as if the charged particles were absent.
At sufficiently early times the neutral flow can be calculated by solving
Euler's equations.
The flow of the charged fluid is driven by and slaved to
the neutral flow by friction.

(iii) We calculated the charged flow for special cases where
the linearized equations of motion are accurate, and carried out an 
extensive analysis of linear MHD waves driven by friction.
The physics of driven waves is embodied in certain Green
functions which describe wave propagation on short time scales,
ambipolar diffusion on long time scales, and transitional behavior
at intermediate times.

(iv) As an illustrative example, we simulated
the collision of two identical clouds with $\Delta v=20$\,\kms.
The simulation incorporated a few unrealistic approximations.
We have argued that the results are qualitatively correct
and illustrate the basic physics.
Realistic solutions will be presented elsewhere.

(v) We found that the formation of a multifluid shock wave
proceeds through two initial phases: an ``ion-electron transient''
and a ``diffusion phase.''
In the former, the cloud-cloud collision produces J shocks in the
neutral fluid which drive wavelike transients into the charged fluid.
The transients quickly evolve into magnetic precursors on the J shocks,
wherein the ions undergo force free motion and the magnetic field
grows steadily in time.
In the diffusion phase, the charged flow continues to evolve in
what appears to be self similar fashion.
The magnetic precursors do not become steady at the largest times we
can study, which are determined by the onset of nonlinearity.

\section*{Acknowledgments}

This work was supported by the New York Center for
Studies on the Origins of Life (NSCORT) and the Department of Physics,
Applied Physics, and Astronomy at Rensselaer Polytechnic Institute,
under NASA grant NAG 5-7589.
We thank the referee, Pierre Lesaffre, for a careful reading of the
manuscript and for comments that improved the presentation.


\clearpage
\appendix
\section{Fourier Integrals for the Green Functions}
\label{app-green}

In order to find the Green functions it is first necessary to
calculate \Gp\xt\ and \Gm\xt.
This involves the evaluation of certain Fourier integrals which
are almost identical to integrals that appear in the
Green functions for the Klein-Gordon equation.
It turns out that ``our'' integrals can be evaluated by exactly
the same technique used in the Klein-Gordon problem.
In \S\S\ref{sec-Gp}--\ref{sec-Gm}
we closely follow the technique of Wyld (1999, see pp.~573ff).

\subsection{Fourier Integral for $G_+$}
\label{sec-Gp}

\begin{figure}
\includegraphics[width=84mm]{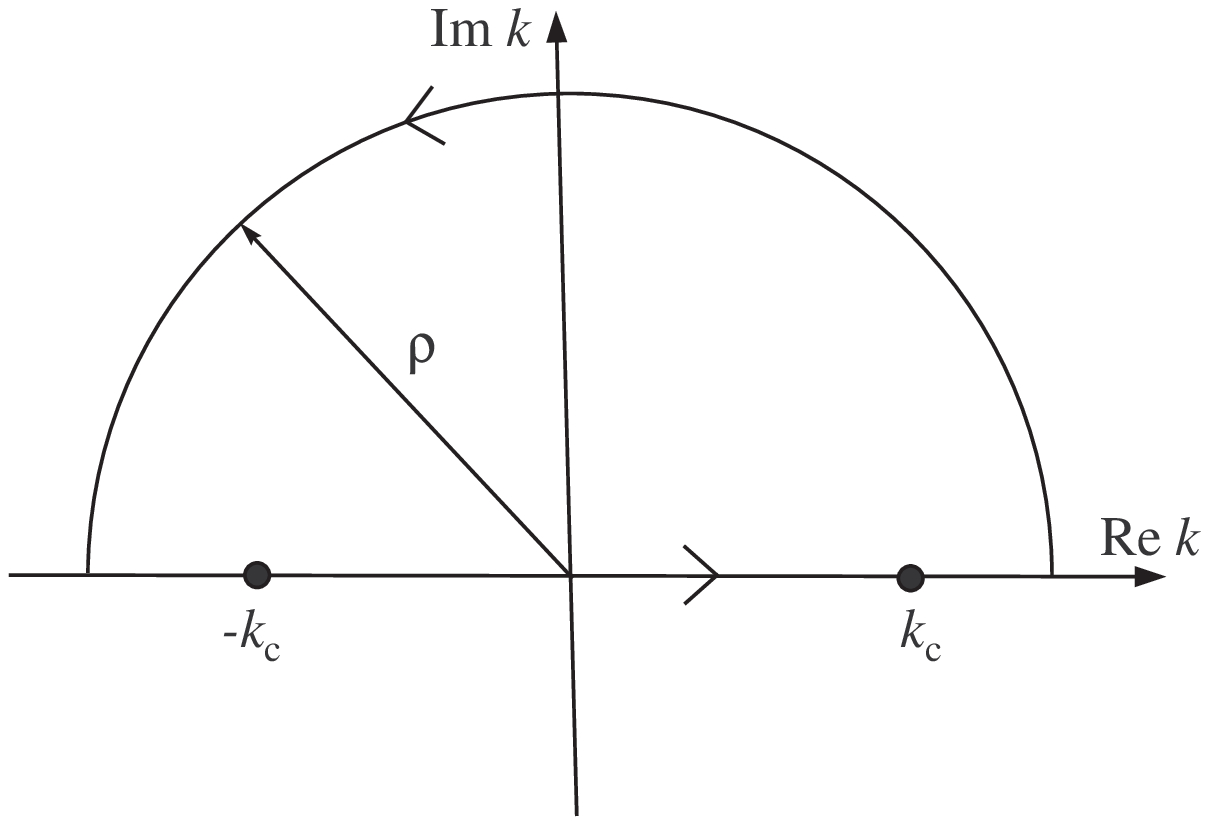}
\caption{Contour used to evaluate \Gp\xt\ for $x>t$
and \Gm\xt\ for $x>-t$. See text.}
\label{fig-contourup}
\end{figure}

\begin{figure}
\includegraphics[width=84mm]{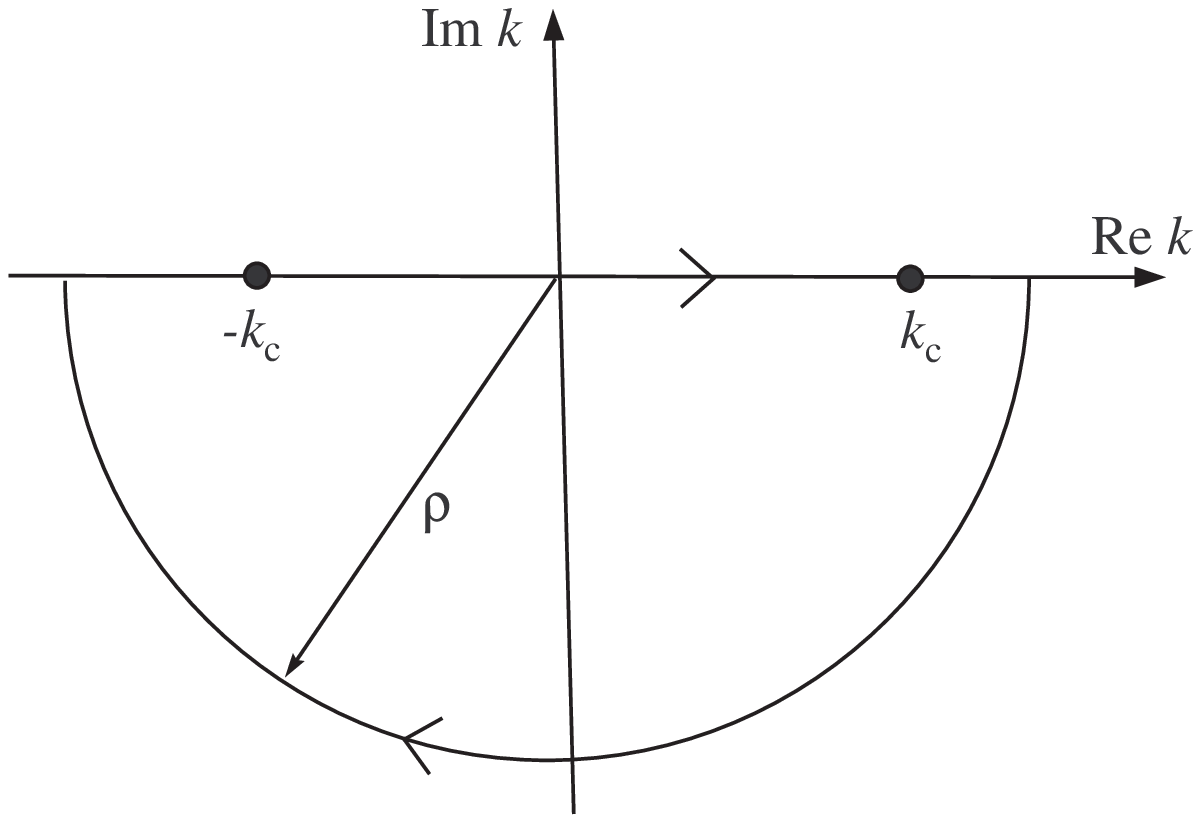}
\caption{Contour used to evaluate \Gp\xt\ for $x<t$
and \Gm\xt\ for $x<-t$. See text.}
\label{fig-contourdown}
\end{figure}

The function
\be
\Gp\xt = \frac{\imath}{4\pi}\, e^{-t/2}\,
\int_{-\infty}^{+\infty} 
\frac{dk\,e^{\imath\left(kx-Rt\right)}}{R}
\label{eq-Gpdef}
\ee
can be evaluated by contour integration as follows.
Noting that $R\rightarrow k$ as $\left|k\right|\rightarrow\infty$,
we see that
\be
e^{\imath\left(kx-Rt\right)} \longrightarrow
e^{\imath k(x-t)}   ~~~~{\rm as}~ \modk\,\longrightarrow\infty.
\ee
In the limit $\modk\goesto\infty$, the
integrand in eq.~(\ref{eq-Gpdef})
%
%
%
goes to zero exponentially in the upper $k$ plane
if $x-t>0$ and in the lower $k$ plane if $x-t<0$.
It follows that we can find \Gp\xt\ by integrating around the semicircular
contour in Fig.~\ref{fig-contourup} if $x>t$ or the semicirle in
Fig.~\ref{fig-contourdown} if $x<t$,
and then taking the limit of infinite radius, $\rho$.
Because of the branch cut in $R$ it is important to
remember that the horizontal
part of each semicircle lies ``just above'' the real $k$ axis
(see \S{\ref{sec-dispersion}).

\begin{figure}
\includegraphics[width=84mm]{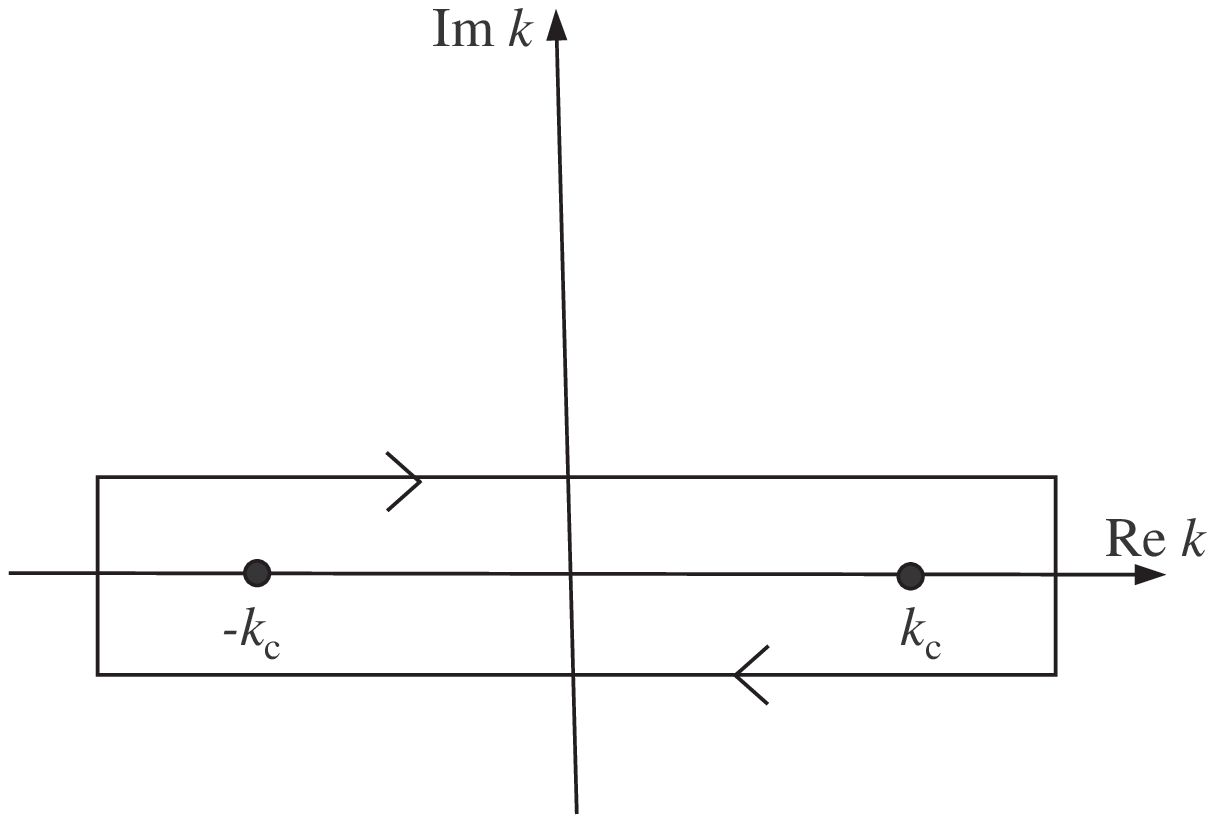}
\caption{Integrals around the closed contours in Fig.~\ref{fig-contourdown}
and the one above are identical according to Cauchy's Theorem
(after Wyld 1999, see p.~577).}
\label{fig-contoursq}
\end{figure}

Consider first the case $x>t$ (Fig.~{\ref{fig-contourup}).
Since there are no singularities in the upper $k$ plane, we find 
immediately that
\be
\Gp\xt = 0 ~~~~{\rm if}~x>t.
\ee
Now consider the case $x<t$ (Fig.~{\ref{fig-contourdown}).
By Cauchy's Theorem, the contour in Fig.~\ref{fig-contourdown}
can be deformed into the one in Fig.~{\ref{fig-contoursq} without
changing the integral.
If we simultaneously change variables from $k$ to $z$, where
\be
k \equiv \kc\,\cosh z,
\label{eq-zdef}
\ee
then expression (\ref{eq-Gpdef}) becomes
\be
\Gp\xt = \frac{\imath}{4\pi}\, e^{-t/2}\, \int_{2\pi\imath}^0\ dz\,
\exp\left[\imath \kc\left(x\cosh z - t \sinh z\right)\right],
\label{eq-Gpintz1}
\ee
where the contour is now a line segment, as indicated by the limits
of integration.
To proceed it is necessary to distinguish whether $\modx < t$
or $\modx > t$. 

Suppose first that $x<t$ and $\modx<t$ (i.e., $-t<x<t$).
Introduce the parameter $\theta$ defined implicitly by
\be
\cosh\theta \equiv \frac{t}{\sqrt{t^2-x^2}}.
\ee
Then
\be
x = \xi\sinh\theta,
\ee
\be
t = \xi\cosh\theta,
\ee
where
\be
\xi(x,t) \equiv \sqrt{t^2-x^2}.
\ee
After eliminating $x$ and $t$ in favor of $\xi$ and $\theta$
in eq.~(\ref{eq-Gpintz1}), 
and using some identities for hyperbolic functions,
we find that
\be
\Gp\xt = \frac{\imath}{4\pi}\, e^{-t/2}\, \int_{2\pi\imath}^0\ dz\,
\exp\left[\imath \kc \xi \sinh\left(\theta-z\right)\right].
\label{eq-Gpintz2}
\ee
Changing the variable of integration again from
$z$ to $\zprm\equiv z-\theta$ yields
\be
\Gp\xt = \frac{\imath}{4\pi}\, e^{-t/2}\, 
\int_{2\pi\imath-\theta}^{-\theta}\ d\zprm\,
\exp\left(-\imath \kc \xi \sinh\zprm\right).
\label{eq-Gpintzp1}
\ee

\begin{figure}
\includegraphics[width=84mm]{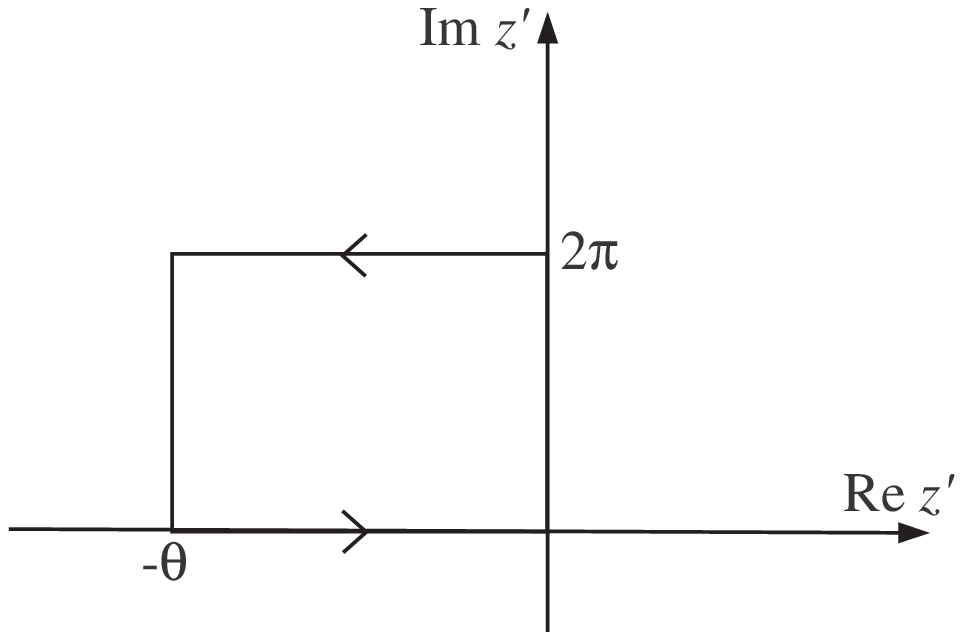}
\caption{Contour used to simplify expression (\ref{eq-Gpintzp1})
(after Wyld (1999, see p.~579).}
\label{fig-contourzpsq}
\end{figure}

The contour for this integral is the left side of the rectangle
in Fig.~\ref{fig-contourzpsq}.
Now the integral around the whole rectangle is zero because
the integrand has no singularities inside the contour.
Further, the integrals along the top and bottom sides
cancel one another because $\sinh(x+2\pi\imath)=\sinh x$. 
It follows that
\be
\Gp\xt = -\frac{\imath}{4\pi}\, e^{-t/2}\, 
\int_0^{2\pi\imath}\ d\zprm\,
\exp\left(-\imath \kc \xi \sinh\zprm\right).
\label{eq-Gpintzp2}
\ee
If we make one more change of variables such that
$\zprm=\imath (y+\pi/2)$, we find that
\be
\Gp\xt = \frac{1}{4\pi}\, e^{-t/2}\, 
\int_{-\pi/2}^{3\pi/2}\ dy\,
\exp\left(\kc \xi \cos y\right).
\label{eq-Gpinty}
\ee
Now the modified Bessel function of order zero has the integral
representation
\be
I_0(x) = \frac{1}{\pi}\,\int_0^{\pi}\ dy\,
\exp\left(\pm x\cos y\right).
\label{eq-I0def}
\ee
Comparing expressions (\ref{eq-Gpinty}) and (\ref{eq-I0def}),
and noting the periodicity of the integrand in the former,
we finally conclude that
\be
\Gp\xt = \frac{1}{2}\,e^{-t/2}\,I_0\left(\kc\xi\right)
~~~{\rm if}~ -t<x<t.
\ee

Next suppose that $x<t$ and $\modx>t$ (i.e., $x<-t$).
Then $x=-\modx$ and eq.~(\ref{eq-Gpintz1}) can be written in the form
\be
\Gp\xt = \frac{\imath}{4\pi}\, e^{-t/2}\, \int_{2\pi\imath}^0\ dz\,
\exp\left[-\imath \kc\left(\modx\cosh z +t \sinh z\right)\right].
\label{eq-Gpintz3}
\ee
Now define $\psi$ by
\be
\cosh\psi \equiv \frac{\left|x\right|}{\sqrt{x^2-t^2}},
\ee
so that
\be
\modx = \zeta\cosh\psi
\ee
and
\be
t = \zeta\sinh\psi,
\ee
where
\be
\zeta(x,t) \equiv \sqrt{x^2-t^2}.
\ee
Eliminating \xt\ in favor of $\zeta$ and $\psi$ changes
eq.~(\ref{eq-Gpintz3}) to
\be
\Gp\xt = \frac{\imath}{4\pi}\, e^{-t/2}\,
\int_{2\pi\imath}^0\ dz\,
\exp\left[-\imath \kc \zeta\,\cosh\left(z+\psi\right)\right].
\label{eq-Gpintz4}
\ee
Changing variables to $\zprm=z+\psi$ changes (\ref{eq-Gpintz4}) to
\be
\Gp\xt = \frac{\imath}{4\pi}\, e^{-t/2}\, 
\int_{2\pi\imath+\psi}^{\theta}\ d\zprm\,
\exp\left(-\imath \kc \zeta\,\cosh\zprm\right)
\label{eq-Gpintzp5}
\ee
and considerations analogous to the ones leading from expression
(\ref{eq-Gpintzp1}) to (\ref{eq-Gpintzp2}) give
\be
\Gp\xt = -\frac{\imath}{4\pi}\, e^{-t/2}\, 
\int_0^{2\pi\imath}\ d\zprm\,
\exp\left(-\imath \kc \zeta\,\cosh\zprm\right).
\label{eq-Gpintzp6}
\ee
The final transformation $\zprm=\imath y$ yields
\be
\Gp\xt = \frac{1}{4\pi}\, e^{-t/2}\, 
\int_0^{2\pi}\ dy\,
\exp\left(-\imath \kc \cos y\right).
\label{eq-Gpintyy}
\ee
Noting that the ordinary Bessel function of order zero is
\be
J_0(x) = \frac{1}{2\pi}\,\int_0^{2\pi}\ dy\, 
\exp\left(\pm\imath x\cos y\right),
\ee
we conclude that
\be
\Gp\xt = \frac{1}{2}\,e^{-t/2}\,J_0\left(\kc\zeta\right) ~~~{\rm if} ~x<-t.
\ee
To summarize:
\be
\Gp\xt = \left\{
\begin{array}{ll}
\frac{1}{2}\,e^{-t/2}\,J_0\left(\kc\zeta\right) & {\rm if}~x<-t \\
 & \\
\frac{1}{2}\,e^{-t/2}\,I_0\left(\kc\xi\right) & {\rm if}~-t<x<t \\
 & \\
0 & {\rm if}~x>t \\
\end{array}
\right.
\label{eq-Gpsum}
\ee

\subsection{Fourier Integral for $G_-$}
\label{sec-Gm}

This calculation is very similar
to the preceding one; details are included for the morbidly curious.
We need to evaluate
\be
\Gm\xt = \frac{\imath}{4\pi}\, e^{-t/2}\,
\int_{-\infty}^{+\infty} 
\frac{dk\,e^{\imath\left(kx+Rt\right)}}{R}.
\label{eq-Gmdef}
\ee
Now the exponential factor behaves like
\be
e^{\imath\left(kx+Rt\right)} \longrightarrow
e^{\imath k(x+t)}   ~~~~{\rm as}~ \modk\,\longrightarrow\infty,
\ee
so we use the contour in Fig.~\ref{fig-contourup} if $x>-t$
and the one in Fig.~\ref{fig-contourdown} if $x<-t$.
Since there are no singularities in the upper $k$ plane we find
immediately that
\be
\Gm\xt = 0 ~~~{\rm if}~x>-t.
\ee
If $x<-t$ we change variables from $k$ to $z$ (cf.~eq~{\ref{eq-zdef})
and deform the contour to the one in Fig.~\ref{fig-contoursq}
with the result
\be
\Gm\xt = \frac{\imath}{4\pi}\, e^{-t/2}\,
\int_{2\pi\imath}^0\ dz\,
\exp\left[-\imath \kc \zeta\,\cosh\left(z-\theta\right)\right].
\label{eq-Gmintz}
\ee
Comparing the integrals in eq.~(\ref{eq-Gpintz3}) and (\ref{eq-Gmintz}),
and noting that the former gives a result which is independent of $\theta$,
we see that
\be
\Gm\xt = \frac{1}{2}\,e^{-t/2}\,J_0\left(\kc\zeta\right) ~~~{\rm if} x<-t.
\ee
To summarize:
\be
\Gm\xt = \left\{
\begin{array}{ll}
\frac{1}{2}\,e^{-t/2}\,J_0\left(\kc\zeta\right) & {\rm if}~x<-t \\
 & \\
0 & {\rm if}~-t<x<t \\
 & \\
0 & {\rm if}~x>t \\
\end{array}
\right..
\label{eq-Gmsum}
\ee

\section{Asymptotic Behavior of the Green Functions}
\label{app-asympt}

For times $\ltsim 1$ the Green functions exhibit complex behavior
including aspects of both wave propagation and diffusion
(Fig.~\ref{fig-greeng}--\ref{fig-gamma}).
However the properties of the dispersion relation (\S\ref{sec-dispersion})
suggest that diffusion dominates at large times, and the
Green functions should behave accordingly.
We now show that this is indeed the case by calculating the Green
functions for $t\gg 1$.

First consider the integral for $G_+\xt$ in eq.~(\ref{eq-Gpdef}).
The time dependent part of the integrand is
\be
e^{-\imath\omega_+ t} = e^{-t/2}\,e^{-\imath Rt}
\ee
and this behaves differently for large and small wave numbers.
Since $R(k)$ is real for $\modk>\kc$, we can neglect
contributions to the integral from $k>\kc=1/2$.
For $\modk<1/2$ the definition of $R$ [eq.~(\ref{eq-Rbranch})] says
\be
R(k) = \frac{\imath}{2}\sqrt{1-4k^2} ~~~~(\modk<1/2)
\ee
so
\be
e^{-\imath\omega_+ t} =
\exp\left\{-\frac{t}{2}\left[1-\left(1-4k^2\right)^{1/2}\right]\right\}
~~~~(\modk<1/2).
\ee
As expected, the integral is dominated by contributions
from $k \ll 1$ (i.e., long wavelengths) at large times.
Setting
\be
e^{-\imath\omega_+ t} \approx e^{-k^2t}
\ee
and
\be
R \approx -\imath/2
\ee
inside the integral sign, we find
\be
G_+\xt \approx\frac{1}{2}\,\int_{-\infty}^{+\infty}\ dk\,
\exp\left(-k^2t+\imath kx\right)
~~~~(t\gg 1).
\ee
The integral is easily evaluated to give
\be
G_+\xt \approx \frac{1}{\sqrt{4\pi t}}\,
e^{-\frac{x^2}{4t}} ~~~~{\rm if}~t \gg 1.
\ee

Next consider the integral in expression (\ref{eq-Gmdef}) for $G_-\xt$.
Now the time dependent factor in the integrand is
\be
e^{-\imath\omega_- t} = e^{-t/2}\,e^{+\imath Rt}.
\ee
Once again contributions to the integral from $\modk > 1/2$
can be neglected.
For $\modk<1/2$, the definition of $R(k)$ implies
\be
e^{-\imath\omega_- t} = 
\exp\left\{-\frac{t}{2}\left[1+\left(1-4k^2\right)^{1/2}\right]\right\}
~~~~(\modk<1/2).
\ee
Since this factor decays faster than $\exp\left(-t/2\right)$ for
all wave numbers, we find
\be
G_-\xt \approx 0 ~~~~{\rm if}~t \gg 1
\ee
and taking the difference of $G_+$ and $G_-$ gives
\be
G\xt \approx \frac{1}{\sqrt{4\pi t}}\,e^{-\frac{x^2}{4t}}
~~~~{\rm if}~t \gg 1.
\ee
Taking partial derivatives gives the other Green functions:
\be
\Gdot\xt\approx \Lambda\xt \approx 
\left[\left(\frac{x}{2t}\right)^2-\frac{1}{2t}\right]\,G\xt
~~~~{\rm if}~t \gg 1,
\ee
and
\be
\Gprm\xt\approx \Gamma\xt \approx 
-\left(\frac{x}{2t}\right)\,G\xt
~~~~{\rm if}~t \gg 1.
\ee
Notice that
\be
\left|\Gdot\xt\right| \ltsim t^{-1}\,\left|G\xt\right|
\ee
and
\be
\left|\Gprm\xt\right| \ltsim t^{-1/2}\,\left|G\xt\right|
\ee
for large times.

\end{document}